\documentstyle[psfig,times]{l-aa}

\begin{document}

   \thesaurus{03         
              (11.03.2;  
               11.05.2;  
               11.09.1;  
               11.09.2;  
               11.19.3;  
               11.19.4)} 

\title{Globular clusters in the blue compact galaxy ESO~338-IG04 (Tol~1924-416), as tracers of the star formation history\thanks{Based on observations with the NASA/ESA {\it Hubble Space Telescope}, obtained at the Space Telescope Science Institute, which is operated by the association of Universities for Research in Astronomy, Inc., under NASA contract NAS5-26555.}}

  \subtitle{Results from HST/WFPC2 observations}

   \author{ G\"oran \"Ostlin
\and 
Nils Bergvall
 \and
Jari R\"onnback        }

   \offprints{G\"oran \"Ostlin, east@astro.uu.se}

   \institute{Astronomiska observatoriet\\
              Box 515\\
              S-75120 Uppsala\\
              Sweden }

   \date{Accepted 16 March 1998}

   \maketitle



\markboth{\"Ostlin et al., Young and old globular clusters in ESO~338-IG04}{}

   \begin{abstract}

Multicolour images of the starbursting metal poor blue compact galaxy ESO~338-IG04 have been obtained with the {\it Wide Field Planetary Camera 2} on board the {\it Hubble Space Telescope}. In the images we find numerous point-like sources concentrated towards the main body of the galaxy, which we identify as globular cluster candidates. We show that these objects are physically associated with the galaxy and that they are  spatially extended. Given their high intrinsic luminosities, these objects cannot be individual stars.  Using photometric evolution models we show that the objects constitute a rich population of  massive star clusters with ages ranging from a few Myr to $\sim$ 10 Gyr, and masses ranging from  $10^4$ to more than $10^7 {\cal M_{\odot}}$. There are peaks in the age distribution of the clusters: one with objects $\le30$ Myr, one at $\sim 100$~ Myr, one at  $\sim 600$~ Myr, one to two at $2.5-5$ Gyr and one at $\sim10$ Gyr. The youngest objects are predominantly found in the crowded starburst region. They have  properties which agree with what is expected for young globular clusters, although it cannot be excluded that some of them may be dissolved or disrupted.  For objects older than a few times 10 Myr, the only plausible explanation is that these are globular clusters. The galaxy presently appears to be involved in a merger, which is the probable cause of the present globular cluster formation. The presence of a numerous intermediate age (2.5 to 5 Gyr) population of globular clusters, suggests that a previous merger might have occurred. As the starburst fades, this galaxy will become very rich in globular clusters. Transforming all objects to an age comparable to that of Milky Way globular clusters  reveals a luminosity function similar to the Galactic. We suggest that this galaxy is the result of a merger between a dwarf  elliptical and a gas rich dwarf. The possibility of dating the globular clusters offers an efficient way of studying the history of violent star formation in this and similar galaxies. 
   
      \keywords{galaxies: compact -- galaxies:  starburst -- galaxies: star clusters -- galaxies: evolution  -- galaxies: interactions -- galaxies: individual: ESO~338-IG04, Tol~1924-416

               }
   \end{abstract}

%

\section{Introduction}
Globular clusters (GCs) are generally old stellar systems and are believed to be the first objects to form in the process of the formation of a galaxy. This is supported by age estimates from observations of GCs in the Milky Way and other nearby galaxies. The Large Magellanic Cloud (LMC) is however known to host several young blue "populous" clusters, which could be young globular clusters. In the recent years blue globular cluster candidates have been found in some interacting/merging galaxies such as NGC~3597 (Lutz 1990, Holtzman et al. 1996), NGC~7252 (Whitmore et al. 1993), "The Antennae" (Whitmore \& Schweizer 1995) and NGC~3921 (Schweizer et al. 1996). This has strengthened the idea  that GCs can form not only when galaxies form, but also when galaxies are reformed in the process of galaxy mergers (Schweizer 1986, Ashman \& Zepf 1992, Whitmore 1996). This could possibly also circumvent the problem, faced by the idea that elliptical galaxies form by the merging of late type galaxies, that ellipticals have higher specific frequencies of globular clusters (van den Bergh 1984, 1994). Understanding how and when GCs can form will thus not only help us in understanding these systems, but also aid in understanding galaxy formation and evolution.

Studies of actively star-forming galaxies, e.g. Henize 2-10 (Conti et al. 1994), NGC~1244 (Hunter et al. 1994), NGC~1705 (O'Connell et al. 1994), NGC~1569 (De Marchi et al. 1997) and M82 (O'Connell 1995) have revealed the presence of star clusters with luminosities comparable to R126, the central cluster of 30 Doradus in the LMC, and higher. 
Meurer et al.(1995) used the HST for an ultraviolet study of nine starburst galaxies and  found bright "super star clusters" in all. These clusters are generally bluer than similar objects found in  interacting galaxies, and are preferentially found in the central regions of the starbursts. It has been proposed (e.g. Conti et al. 1994) that these blue objects might be forming globular clusters. This is however still an open question since it is not clear if these systems will survive as gravitationally bound systems. Uncertainties in the value of the (universal ?) stellar initial mass function (IMF) makes the mass estimates, mainly based on ultraviolet  data, highly uncertain, especially since some studies only includes one spectral bandpass. Many of these galaxies appear to be involved in some form of interaction.

One could interpret  these observations as evidence that massive star clusters can form in regions of very active star formation, such as giant extragalactic HII-regions (GEHRs)  like 30 Doradus. Kennicutt and Chu (1988)  made a statistical investigation of data on extragalactic blue populous clusters and giant HII-regions. They concluded that the young clusters in LMC are not luminous enough to evolve into globular clusters comparable to the massive galactic ones, and that far from all  GEHRs will produce young populous clusters that could become GCs.

  Blue compact galaxies (BCGs) are characterised by their blue colours ($B-V \le 0.5$), strong nebular emission lines, indicative of the formation of relatively hot (massive) stars, and low chemical abundances. The derived star formation rates could, considering the gas supply, only be sustained for a small fraction of a Hubble time. This together with the low metallicities (with IZw18 being the extreme in this sense)  once lead to the idea that BCGs might be genuinely young objects now experiencing their first star formation  epoch (Sargent and Searle 1970). Now, most BCGs are believed to be old, experiencing recurrent bursts of star formation intervened by long quiescent periods. Still we do not yet understand what triggers these bursts of star formation. One possible trigger mechanism is tidal or direct interactions with companion galaxies or gas clouds.

In this paper we will present conclusive evidence for young and old globular clusters in the blue compact galaxy ESO~338-IG04. Section 2 describes the observations and Sect. 3 describes how photometry was performed on the globular cluster candidates. In Sect. 4 we show that the objects are physically associated with the galaxy and that they are spatially resolved. In Sect. 5 we discuss how the observed photometric properties can be interpreted in terms of age and mass of the objects. Section 6 includes a further discussion on the nature of the cluster candidates, and Sect. 7 contains the conclusions.

\subsection{General properties of the target galaxy}

ESO~338-IG04, also known as Tol~1924-416, resides at a distance of 37.5 Mpc ($v_{hel}=2813$ km~s$^{-1}$, $H_0= 75$ km s$^{-1}$ Mpc$^{-1}$; this value will be used throughout the rest of the paper) at the celestial coordinates $\alpha_{1950}=19^{\rm h}~24^{\rm m}~30^{\rm s}~ \delta_{1950} = -41\degr~34\arcmin~00\arcsec$. It  is intrinsically bright ($M_V=-19.3$) and blue ($B-V=0.4$), (Bergvall and \"Ostlin 1998). The oxygen abundance is $12 \%$ of the solar value (Bergvall, 1985). It's physical size is small, 
$8.5 \times 4.5$ 
kpc measured at $\mu _V=25 {\rm mag}~{\rm arcsec}^{-2}$, the non dwarfish luminosity being due to the active star burst. It has an integrated HI-mass of   $3\times10^9 {\cal M_{\odot}}$ (\"Ostlin et al. 1997b). The optical velocity field of ESO~338-IG04 show rotation aligned with it's apparent major axis, though with several large scale irregular features, and it is most easily understood as a merger between two galaxies or a galaxy and a gas cloud (\"Ostlin et al. 1997a). This is supported by the optical tail, extending towards the east. It has a spectroscopically confirmed companion galaxy (Bergvall unpublished; \"Ostlin et al. 1997a) which lies at a projected distance of 70 kpc, see Fig. 3. The companion is a somewhat fainter ($M_V = -17.9$), star forming galaxy which shows regular rotation (\"Ostlin et al. 1997a).  Bergvall noticed, in ground based images, an apparent concentration of faint blobs around the galaxy, which he interpreted as globular cluster candidates. We have re-observed  this galaxy with the HST and will in the subsequent sections show that there is substantial evidence that  this galaxy has formed massive globular clusters at several epochs.

\begin{figure}
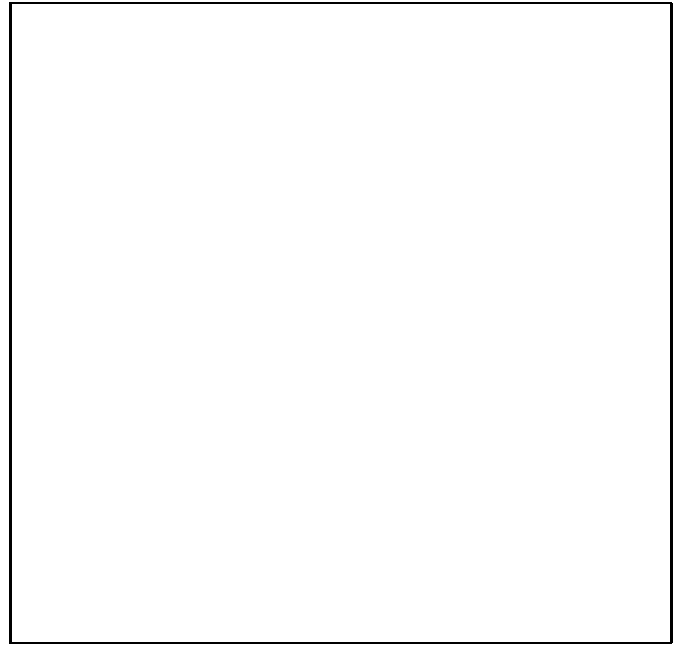

\picplace{8.5cm}
\caption[]{A Digitized Sky Survey (DSS) image (obtained with $Skyview$) showing the target galaxy and its companion. North is up and  east is left. The total angular size of the field is ~$9\arcmin \times 9\arcmin$. ESO~338-IG04 is at the upper left, the companion at the lower right. }
\end{figure}

In July 1997, we obtained spectra of a few of the brightest globular cluster candidates, using the ESO New Technology Telescope (NTT), at La Silla, Chile. The results from this spectroscopic study will be presented in a future publication.

\section{Observations and reductions} The field of ESO~338-IG04 was observed with the planetary camera (PC) aperture (centred on $\alpha_{2000}=19^{\rm h}~28^{\rm m}~00^{\rm s}.23$, ~$\delta_{2000} -41\degr~34\arcmin~21\arcsec.2$, with the HST V3-axis at position angle 70.0 degrees) of WFPC2 onboard Hubble Space Telescope on April 29, 1996. Exposures were taken in the F218W(1800s), F336W(900s), F439W(800s), F555W(260s) and F814W(520s) filters. Each exposure was split in two to facilitate removal of cosmic rays.
The work presented in this paper is based on the F336W, F439W, F555W and F814W data. The F218W data is underexposed and will be used only for the brightest super star clusters in the central starburst component, which will be discussed in a forthcoming paper. The pixel scale of the PC CCD is 0.0455 arcseconds, which at a distance of 37.5 Mpc corresponds to 8.27 pc.

\begin{figure*}
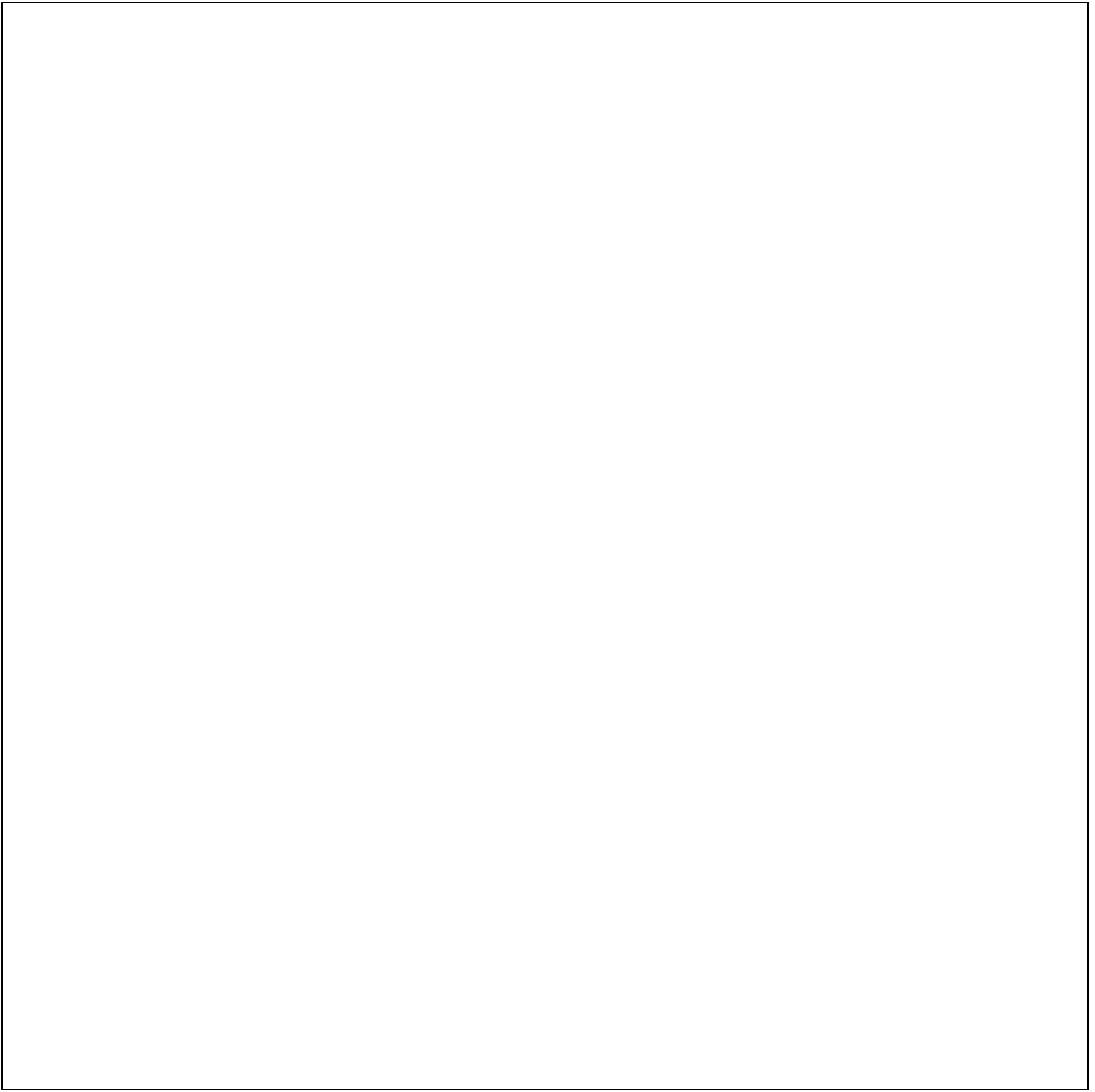

\picplace{18cm}
\caption[]{The F814W image of the   planetary camera (PC) field centred on ESO~338-IG04. The image has been trimmed slightly, not to display the dirty regions of the PC chip on the sides next to the WFC chips. The displayed field is approximately $33\arcsec \times 33\arcsec$, which at the assumed distance of 37.5 Mpc corresponds to $6 \times 6$ kpc. The orientation is indicated by the arrow: north is up-left (arrow head), east is down-left. The length of the arrow is 18.2\arcsec, or $\sim$ 60 pc. The intensity scale is logarithmic. The globular cluster candidates are visible outside the central starburst component, and there is a clear concentration of faint point-like objects around the galaxy.  The isophote where the centre, in this brightness scaling, merges into one black blob, approximately coincides with the division line between the outer and inner sample, defined in the text. The bright object $\sim5\arcsec$ west from the field centre is a foreground star}
\end{figure*}

Data produced with the standard pipeline processing was used. Additional reductions was performed with the STSDAS package in IRAF. Additional hot pixels were examined using the {\tt warmpix} task. Images cleaned from cosmic rays were produced using the {\tt crrej} task and the few remaining bad pixels were removed with the {\tt cosmicrays} task. The F336W data have been corrected for "UV contamination" (a gradual loss of UV throughput in between the monthly heating of the  CCDs) using the quoted values in the HST data handbook (Table 41.2). We used the photometric zero-points for Vega (VEGAMAG) given in the HST Data Handbook, version 2, (Table 41.1), and will use the notation $m_{336}$, $m_{439}$, $m_{555}$ and $m_{814}$ for the magnitudes in this system, sometimes $u$, $b$, $v$, and $i$ may be used, e.g. for colour indices, $(v-i)$ etc. These zero-points give colours that agree with those of the standard $U~B~V~R~I~$ Kron-Cousins (KC) system for A0V stars. For stars with spectral type differing a lot from A0V and for composite stellar systems (e.g. galaxies), the colours in the HST-Vega system may differ appreciably from the KC colours (WFPC2 instrument handbook, version 4, Tab 8.1, see also Holtzman et al. 1995b).  For most objects our resulting photometry is reasonably close to the standard KC  $U$ (F336W) $B$ (F439W), $V$ (F555W) and $I$ (F814W) filter system. However, as is shown in section 5.1, our results are not sensitive to the similarity with the KC-system, since we used the  actual filter profiles and model spectra to interpret the observed colours.  For comparison with other published data on extragalactic globular cluster systems it is however convenient to use a photometric system close to the standard.

\begin{figure*}
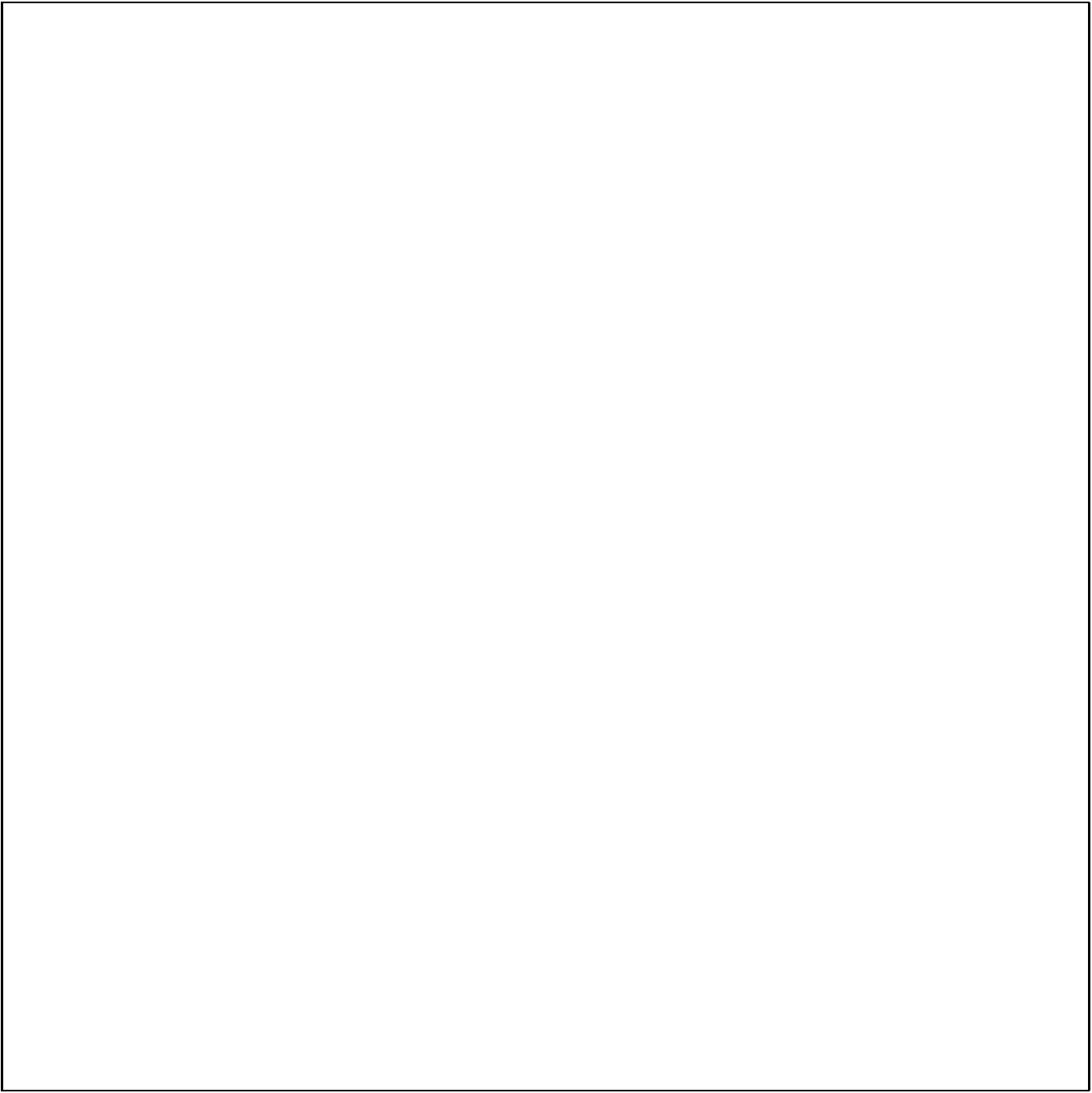

\picplace{18cm}
\caption[]{A magnification of the $11\arcsec \times 11\arcsec$ starburst region of ESO~338-IG04 in filter F814W. The orientation is the same as in Fig. 1. The image has been scaled in order to enhance the contrast  to show the location of bright point-like objects in the central parts. Some of these objects were identified in the ultraviolet study by Meurer et al. (1995).  }
\end{figure*}

The reduced images show a bright "chaotic" starburst region with several very bright knots and filamentary  emission, superimposed on a redder ($b-v \approx 0.9$) component with lower surface brightness ($\mu_{0, v} \sim 20~ {\rm mag~ arcsec}^{-2}$, the extrapolated central surface brightness obtained by fitting an inclination corrected disc between 1.5 and 2.5 kpc of the underlying component). The starburst region is surrounded by a system of point-like objects clearly concentrated towards the centre of the galaxy. Figure 2 shows the reduced F814W image of  ESO~338-IG04, and Fig. 3 a magnification of the central parts. Due to the fairly short exposure times, the faint outer parts of the galaxy are not clearly visible.  This means that most objects found outside the bright central component in the PC field of view are actually within ESO~338-IG04. The faintest parts of the galaxy extend far from the starburst centre, into the wide field CCDs, primary WFC4 and WFC3. In addition to the objects presented in this paper, the WFC CCDs contain a few objects which likely are associated with ESO~338-IG04.

\section{Photometry of globular cluster candidates}

Figures 1 and 2 reveals several point-like objects. It is not obvious how to perform photometry of these
objects. For objects outside the main body, crowding problems are generally small and normal aperture photometry will be adequate. For the sources in the centre, several complications add. First, crowding poses severe difficulties in subtracting the background, and secondly gas emission, both line and continuos, associated with the bright young clusters and the filaments contaminate the broad-band colours and adds to creating an uneven background. 
These two subsets of objects will be referred to as the$~outer~$ and the$~inner~$sample.

To estimate the amount of Milky Way extinction along the line of sight we used the data by Burstein and Heiles (1982) which gives $E(B-V)=0.09$. To  de-redden the data we applied this extinction and the average interstellar extinction curve (Osterbrock, 1989).  N.B., in all figures and tables in this paper, the data have been corrected for this Galactic reddening. In addition there will be internal extinction in the target galaxy and in the cluster candidates themselves; this issue will be discussed in Sect. 5.3. In none of the tables or figures, any correction for internal extinction has been applied to the data.
To convert apparent magnitudes to absolute we used a distance modulus of 32.87 based on the distance 37.5 Mpc and the above correction for Milky Way extinction. This of course assumes all objects to be at the distance of the target galaxy, which as we will see in Sect. 4 is strongly supported.

\subsection{Sources outside the crowded main body, the outer sample}

We have identified all faint point-like sources in the PC images by eye, excluding obvious galaxies and diffuse sources.  For the sources outside the crowded central regions, $\sim 70$ sources brighter than $m_{814} = 24$ (before extinction and aperture correction) were found. To obtain photometry of these we have used the IRAF task {\tt phot} in the {\tt apphot} package. Photometry was  performed at 40 different apertures with radii ranging from 0.2 to 9 pixels. Then a mean growth curve were constructed from the best (high signal-to-noise and well behaved growth curve) objects for each filter. The magnitude of each individual object have been calculated by averaging up to 8 points (fewer for faint objects with noisy growth curve) where the differential growth curve (the difference between the growth curve for the object and the mean growth curve) is flat, normalised to an aperture with a radius of 9 pixels. This gives more accurate results than single aperture photometry since the magnitude is calculated from several points where the profile is well behaved, e.g. image artefacts like remaining hits will have little effect on photometry. Furthermore, as will be shown, the objects are generally spatially resolved; therefore methods using point spread function (PSF) fitting, like DAOPHOT, will not work well. The sky was subtracted
using the median pixel value in a 4 pixels wide annulus starting at a radius of 10 pixels from the object to be measured. Aperture corrections were determined using Table 2a in Holtzman et al. (1995a) and taking into account both the fraction of the stellar flux outside the  aperture radius of 9 pixels and the fraction of the stellar flux falling into the sky annulus. This photometric scheme works well for all objects outside the main body where crowding is not a problem. We have experimented with using smaller apertures and a closer sky annulus and have obtained consistent results. 
It is possible to achieve a higher formal signal-to-noise (S/N), as measured by the photon statistics, by using a smaller aperture. This has the cost that centring errors will be introduced and the aperture corrections will be less accurate, depending on the extent of the object; i.e. the effective S/N will not be higher.

The data have not been corrected for geometric distortion in the PC since the objects are generally found far from the edges of the chip and the induced errors would in all cases be smaller than the photometric uncertainties (maximum a few percent at the corners of the chip, HST Data Handbook). Furthermore, geometric distortion would not alter the colours of objects, on which most of the analysis here is based.  The photometry is presented in Table 1 for the F336W, F439W, F555W and F814W data, where we have used the zero points for Vega.  Figure 4 displays the photometric errors as a function of magnitude for the four different filters. Due to their typical colours, most objects are found in the F814W images, and least in the F336W images.

\begin{table*}
\caption[]{Photometry of objects in the outer sample with $m_{814} < 24 $ . For explanation of column entries, see caption of Table 2.  An asterisk(*) means that the absolute $i$ magnitude is given rather than $v$. Double and triple asterisks indicates that the object is a probable respectively possible, background galaxy/foreground star.} 
\begin{flushleft}
\begin{tabular}{lllllllllllllll}
\noalign{\smallskip}
\hline
\noalign{\smallskip}
Object & x-coord & y-coord & $m_{814}$ & $\sigma_{814}$ & $m_{555}$ & $\sigma_{555}$ & $m_{439}$ & $\sigma_{439}$ & $m_{336}$ & $\sigma_{336}$ & $M_{v}$  & $(v-i)$ & $(b-v)$ & $(u-b)$ \\
\noalign{\smallskip}
\hline\noalign{\smallskip}
 1 & 123.4 &  98.0  &  22.39 &  0.06  & 22.77  & 0.08  & 22.80 &  0.10 &  21.99  & 0.22  & -10.1  &  0.39  &  0.02 &  -0.81  \\
 2 & 187.0 &  89.4 &  23.05 &  0.11  & 23.60  & 0.15  & 23.64 &  0.17 &  24.54  & 1.74  &  -9.3  &  0.55  &  0.04 &   0.89   \\ 
 3 & 250.4 & 121.0 &  22.61 &  0.09  & 23.39  & 0.10  & 23.52 &  0.17 &  22.46  & 0.23  &  -9.5  &  0.78  &  0.13 &  -1.06   \\ 
 4 & 238.0 & 132.0 &  21.58 &  0.03  & 22.04  & 0.04  & 21.98 &  0.05 &  20.99  & 0.07  & -10.8  &  0.47  & -0.06 &  -0.99   \\ 
 5 & 292.0 & 104.0  &  23.47 &  0.12  & 23.86  & 0.17  & 24.17 &  0.29 &  22.84  & 0.33  &  -9.0  &  0.39  &  0.31 &  -1.33   \\  
 6 & 312.6 & 145.3 &  23.40 &  0.11  & 24.48  & 0.31  & 24.36 &  0.47 &  24.69  & 2.37  &  -8.4  &  1.08  & -0.13 &   0.33   \\ 
 7 & 266.6 & 169.0 &  23.77 &  0.19  & 25.17  & 0.47  & -     & -     &   -     & -     &  -7.7  &  1.40  & -     &  -  \\   
 9 & 185.0 & 145.0 &  23.59 &  0.16  & 24.49  & 0.23  & 23.85 &  0.28 &   -     & -     &  -8.4  &  0.90  & -0.64 &  -  \\  
10 & 155.6 & 139.4 &  23.83 &  0.21  & 25.63  & 0.69  & -     & -     &  23.28  & 0.56  &  -7.2  &  1.80  & - &  -  \\  
11 & 103.0 & 212.4 &  22.23 &  0.04  & 23.06  & 0.08  & 23.67 &  0.18 &  23.01  & 0.55  &  -9.8  &  0.82  &  0.62 &  -0.66   \\ 
12 & 171.0 & 191.0 &  23.36 &  0.12  & 24.18  & 0.18  & 24.52 &  0.55 &   -     & -     &  -8.7  &  0.82  &  0.34 &  -  \\  
13$^{***}$ & 140.4 & 399.6 &  22.85 &  0.07  & 24.50  & 0.20  & -     & -     &  23.71  & 0.67  &  -8.4  &  1.65  & - &  -  \\  
14$^{**}$ & 182.0 & 401.2 &  23.62 &  0.14  & -      & -     & -     & -     &  24.27  & 1.11  &  -9.2$^*$ & $>$1.4 & -      &  -      \\  
15 & 252.4 & 410.6 &  23.90 &  0.19  & 25.24  & 0.61  & -     & -     &   -     & -     &  -7.6  &  1.34  & - &  -  \\  
16 & 234.3 & 375.0 &  23.90 &  0.28  & 24.74  & 0.56  & -     & -     &   -     & -     &  -8.1  &  0.84  & - &  -  \\  
17 & 233.1 & 362.1 &  23.74 &  0.18  & 24.20  & 0.22  & 23.63 &  0.23 &  22.06  & 0.17  &  -8.7  &  0.47  & -0.57 &  -1.57   \\ 
18 & 256.0 & 359.0 &  21.76 &  0.07  & 22.77  & 0.12  & 23.30 &  0.22 &  24.07  & 1.39  & -10.1  &  1.02  &  0.52 &   0.77   \\ 
19$^{***}$ & 170.2 & 368.0 &  23.61 &  0.15  & 25.24  & 0.40  & -     & -     &  23.67  & 0.67  &  -7.6  &  1.63  & - &  -  \\  
20 & 184.0 & 334.2 &  23.60 &  0.15  & 23.91  & 0.18  & 24.22 &  0.28 &  22.79  & 0.36  &  -9.0  &  0.31  &  0.30 &  -1.42   \\ 
21 & 185.0 & 309.3 &  21.38 &  0.03  & 22.24  & 0.05  & 22.83 &  0.10 &  22.45  & 0.26  & -10.6  &  0.86  &  0.58 &  -0.38   \\ 
22 & 218.0 & 296.0 &  20.83 &  0.02  & 21.35  & 0.03  & 21.64 &  0.05 &  21.26  & 0.09  & -11.5  &  0.53  &  0.29 &  -0.39   \\ 
24 & 229.0 & 277.2 &  23.06 &  0.12  & 23.72  & 0.23  & 24.17 &  0.42 &  23.73  & 1.66  &  -9.1  &  0.66  &  0.45 &  -0.44   \\ 
25 & 234.8 & 269.5 &  21.93 &  0.05  & 21.28  & 0.06  & 22.17 &  0.07 &  20.82  & 0.07  & -11.6  & -0.65  &  0.89 &  -1.34   \\ 
26 & 217.0 & 255.0 &  22.26 &  0.05  & 22.84  & 0.07  & 23.51 &  0.18 &  23.83  & 0.99  & -10.0  &  0.59  &  0.67 &   0.33   \\ 
27 & 276.7 & 237.0 &  22.87 &  0.09  & 23.86  & 0.20  & 24.37 &  0.36 &   -     & -     &  -9.0  &  0.98  &  0.51 &  -  \\  
28 & 288.7 & 204.9 &  23.35 &  0.11  & 24.16  & 0.26  & 24.22 &  0.39 &  23.35  & 0.55  &  -8.7  &  0.81  &  0.06 &  -0.87   \\ 
29 & 325.0 & 218.0 &  22.25 &  0.05  & 22.92  & 0.08  & 23.24 &  0.15 &  23.82  & 0.81  &  -9.9  &  0.67  &  0.31 &   0.59   \\ 
30 & 407.0 & 155.1 &  22.11 &  0.04  & 22.96  & 0.08  & 23.52 &  0.16 &  23.19  & 0.47  &  -9.9  &  0.86  &  0.56 &  -0.33   \\ 
31 & 398.2 & 234.2 &  23.10 &  0.10  & 23.26  & 0.09  & 23.58 &  0.15 &  22.27  & 0.20  &  -9.6  &  0.16  &  0.33 &  -1.31   \\ 
32$^{***}$ & 367.5 & 298.0 &  23.43 &  0.24  & 25.26  & 0.70  & 24.70 &  0.55 &   -     & -     &  -7.6  &  1.83  & -0.56 &  -  \\  
33 & 417.3 & 303.0 &  23.02 &  0.08  & 24.03  & 0.16  & 25.85 &  1.55 &  23.57  & 0.60  &  -8.8  &  1.02  &  1.82 &  -2.28   \\ 
34 & 387.7 & 356.2 &  19.29 &  0.01  & 20.17  & 0.01  & 20.71 &  0.02 &  20.95  & 0.08  & -12.7  &  0.87  &  0.54 &   0.24   \\ 
39$^{**}$ & 189.0 & 520.0 &  21.44 &  0.03  & 23.49  & 0.11  & 24.66 &  0.40 &  23.33  & 0.79  &  -9.4  &  2.05  &  1.17 &  -1.33   \\ 
40 & 324.8 & 605.6 &  20.59 &  0.01  & 21.39  & 0.02  & 21.86 &  0.04 &  21.78  & 0.15  & -11.5  &  0.79  &  0.47 &  -0.08   \\ 
41 & 374.0 & 599.3 &  22.51 &  0.06  & 23.40  & 0.11  & 23.83 &  0.22 &   -     & -     &  -9.5  &  0.89  &  0.43 &  -  \\  
42 & 369.1 & 524.6 &  20.59 &  0.02  & 21.17  & 0.02  & 21.45 &  0.03 &  21.27  & 0.11  & -11.7  &  0.58  &  0.28 &  -0.17   \\ 
43 & 403.8 & 567.4 &  22.17 &  0.04  & 23.05  & 0.10  & 23.41 &  0.14 &  24.03  & 1.16  &  -9.8  &  0.87  &  0.36 &   0.62   \\ 
44 & 392.0 & 515.8 &  21.88 &  0.04  & 22.79  & 0.07  & 23.05 &  0.11 &   -     & -     & -10.1  &  0.90  &  0.26 &  -  \\  
45$^{**}$ & 437.0 & 503.2 &  22.60 &  0.07  & 24.62  & 0.30  & -     & -     &  24.90  & 2.41  &  -8.2  &  2.02  & - &  -  \\  
46 & 416.0 & 472.0 &  22.90 &  0.10  & 23.77  & 0.15  & 23.90 &  0.23 &  23.88  & 0.86  &  -9.1  &  0.87  &  0.13 &  -0.02   \\ 
47 & 341.4 & 443.0 &  20.75 &  0.02  & 21.38  & 0.03  & 21.84 &  0.05 &  21.53  & 0.13  & -11.5  &  0.63  &  0.46 &  -0.31   \\ 
48 & 374.0 & 437.2 &  21.85 &  0.04  & 22.60  & 0.06  & 23.33 &  0.19 &  23.48  & 0.74  & -10.3  &  0.75  &  0.73 &   0.15   \\ 
49 & 401.7 & 431.7 &  21.27 &  0.03  & 21.69  & 0.03  & 21.75 &  0.05 &  21.08  & 0.08  & -11.2  &  0.43  &  0.06 &  -0.68   \\ 
50 & 464.0 & 447.0 &  23.38 &  0.11  & 23.99  & 0.18  & 23.80 &  0.27 &  22.35  & 0.26  &  -8.9  &  0.61  & -0.20 &  -1.44   \\ 
51 & 481.9 & 667.0 &  22.65 &  0.06  & 23.71  & 0.13  & 24.07 &  0.25 &   -     & -     &  -9.2  &  1.06  &  0.36 &  - \\  
52 & 619.4 & 698.9 &  21.77 &  0.04  & 22.66  & 0.06  & 23.37 &  0.13 &  23.04  & 0.49  & -10.2  &  0.89  &  0.71 &  -0.33   \\ 
53 & 606.7 & 558.2 &  20.60 &  0.02  & 21.32  & 0.02  & 21.70 &  0.04 &  22.52  & 0.28  & -11.6  &  0.72  &  0.38 &   0.82   \\ 
54 & 599.2 & 475.0 &  23.56 &  0.12  & 24.14  & 0.21  & -     & -     &  23.15  & 0.50  &  -8.7  &  0.58  & - &  -  \\  
55 & 450.3 & 399.5 &  23.08 &  0.09  & 23.96  & 0.16  & 24.72 &  0.47 &  23.44  & 0.54  &  -8.9  &  0.88  &  0.76 &  -1.28   \\ 
56 & 438.3 & 337.6 &  23.35 &  0.12  & 24.44  & 0.21  & 24.41 &  0.33 &  24.09  & 1.25  &  -8.4  &  1.09  & -0.03 &  -0.32   \\ 
57 & 438.4 & 228.9 &  22.99 &  0.08  & 23.89  & 0.17  & 24.21 &  0.38 &  23.33  & 0.52  &  -9.0  &  0.90  &  0.32 &  -0.88   \\ 
58 & 544.6 & 306.4 &  21.35 &  0.02  & 22.43  & 0.05  & 23.20 &  0.15 &  23.08  & 0.52  & -10.4  &  1.08  &  0.76 &  -0.12   \\ 
59 & 593.3 & 366.8 &  21.69 &  0.03  & 22.16  & 0.04  & 22.45 &  0.09 &  22.31  & 0.33  & -10.7  &  0.46  &  0.29 &  -0.13   \\ 
60 & 590.4 & 350.0 &  23.48 &  0.12  & 24.03  & 0.18  & 24.35 &  0.31 &   -     & -     &  -8.8  &  0.55  &  0.32 &  -  \\  
61 & 632.9 & 152.3 &  23.56 &  0.14  & 24.73  & 0.32  & -     & -     &   -     & -     &  -8.1  &  1.18  & - &  -  \\  
63$^{***}$ & 630.0 & 429.8 &  23.00 &  0.09  & 24.83  & 0.31  & 24.77 &  0.48 &   -     & -     &  -8.0  &  1.83  & -0.06 &  -  \\  
64$^{**}$ & 714.0 & 469.7 &  23.03 &  0.09  & -      & -     & -     & -     &   -     & -     & -9.8$^*$  & $>$2.0  & - &  -  \\  
66 & 559.4 & 552.2 &  23.75 &  0.16  & 24.26  & 0.27  & 25.29 &  0.78 &   -     & -     &  -8.6  &  0.51  &  1.03 &  -  \\  
68 & 371.7 & 408.0 &  22.75 &  0.11  & 23.34  & 0.14  & 23.53 &  0.23 &  22.10  & 0.20  &  -9.5  &  0.59  &  0.19 &  -1.43   \\ 
69 & 306.3 & 441.0 &  23.88 &  0.26  & 23.17  & 0.10  & 23.93 &  0.23 &  22.52  & 0.27  &  -9.7  & -0.71  &  0.76 &  -1.41   \\ 
90 & 345.0 & 305.1 &  22.61 &  0.14  & 21.68  & 0.07  & 23.13 &  0.21 &  21.63  & 0.17  & -11.2  & -0.93  &  1.45 &  -1.50   \\ 
91 & 292.2 & 246.5 &  23.08 &  0.10  & 23.02  & 0.10  & 22.69 &  0.10 &  21.32  & 0.11  &  -9.8  & -0.05  & -0.34 &  -1.36   \\ 
92 & 295.0 & 230.2 &  22.42 &  0.07  & 21.36  & 0.03  & 22.07 &  0.06 &  20.68  & 0.06  & -11.5  & -1.06  &  0.72 &  -1.39   \\ 
93 & 367.5 & 298.2 &  23.67 &  0.14  & 25.65  & 1.01  & 25.17 &  0.70 &  -  & -  &  -7.2  &  1.98  & -0.47 &   -  \\  
\hline  
\noalign{\smallskip} 
\end{tabular}
\end{flushleft}
\end{table*}

\subsection{Sources in the central main body, the inner sample}

In the main body there are several very bright point-like objects. In general, these are spatially extended, as will be discussed in Sect. 4.4. Photometry is more difficult here and the above approach would fail due to severe crowding problems. Moreover, the central part of the F555W images are severely affected by [OIII]$_{\lambda5007}$ emission. Contamination from gaseous emission is less problematic in the other filters since no bright lines are within the band-passes. The filaments do however also produce continuous emission which may cause problems for photometry in all filters, mainly because they contribute to make the background highly variable on small scales.  Since the objects are not true point sources it is hazardous to apply standard PSF fitting methods like DAOPHOT. A possible approach would be to fit a mean PSF for the intrinsically extended sources and use DAOPHOT with this as input PSF. This does not work in practice since there is no well defined average profile for the objects and the final residuals from the PSF subtraction becomes unacceptably large. 

\begin{figure}
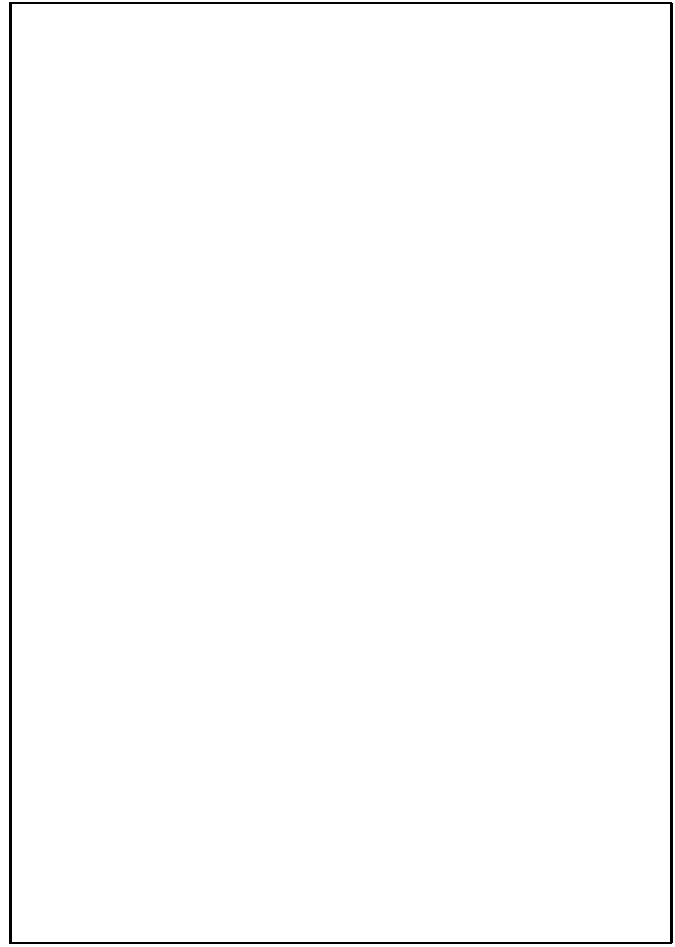

\picplace{12.5cm}
\caption[]{Photometric uncertainties: The left panel shows from the top downwards the F336W, F439W, F555W and F814W data, respectively, for the outer sample. The right panel shows the same for the inner sample.  The x-axis is the magnitude in the pass-band in question, and the y-axis is the 1 sigma photometric uncertainty, as calculated by the IRAF routine {\tt phot}, which is based on the photon statistics. A few sources fall outside the plotted range and are not included, see Table 1 and 2. }
\end{figure}

To extract information on the colours and luminosities of the sources in the centre we instead used simple aperture photometry in small apertures with radius 3 pixels. The sky was subtracted using the median in an annulus from 4 to 6 pixels. The small aperture was selected to minimise problems with neighbouring objects entering the aperture. This has the cost that centring errors may be important and aperture corrections may be inaccurate . The sky annulus was chosen to be close to the objects due the large amplitude of the small scale background brightness fluctuations. 
As for the outer sample, no correction for geometric distortion has been applied.

   Contamination from  [OIII]$_{\lambda5007}$ emission in the F555W band can be serious for these objects (see Fig. 5), therefore more reliable colours of the stellar population may be achieved using the photometry in the F814W, F439W and F336W filters, which are free of strong nebular emission lines. The inner objects are, in  general, bluer than the sources outside the main body. The photometric data for these objects are shown in Table 2 and Fig. 4, and includes the formal photometric uncertainties. Generally the photometry for the inner sample is less certain and the errors are probably larger than the quoted uncertainties, due to centring errors and background subtraction problems. E.g. our experiments show that centring errors may be as large as 0.05 magnitudes. These additional errors are not included in the entries in Table 2. , so for a fair comparison with the outer sample, the errors of the objects in the inner sample should be scaled up with  approximately a factor of two. 

\begin{figure}
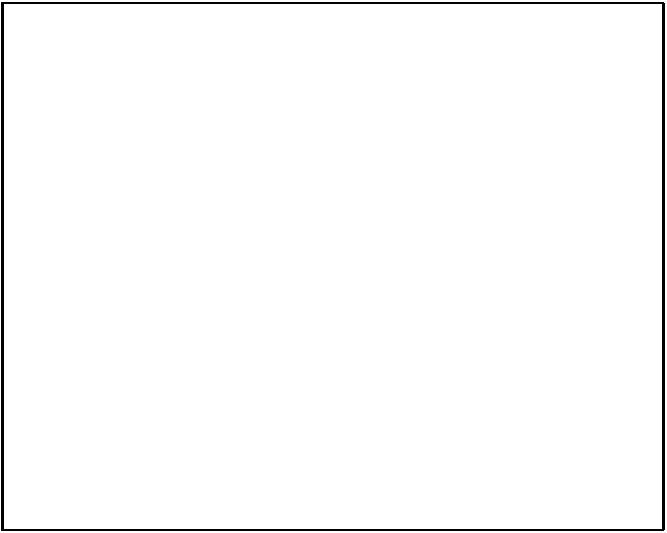

\picplace{7cm}
\caption[]{
Colour-magnitude diagram of all objects in Table 1 and 2. Magnitudes refer to the photometric zero-points for Vega. The objects in the outer sample and inner sample are shown as filled and open circles respectively. The data have been corrected for Galactic reddening. Note that the two subsamples have different magnitude limits; $m_{814}=23.5$ and $m_{814}=24$ for the inner and outer sample, respectively.}
\end{figure}

\begin{table*}
\caption[]{Photometry of objects in the inner sample with $m_{814} < 23.5$. The X and Y coordinates correspond to the pixel numbers in the planetary camera for the observed configuration. For all photometric data the zero-points for Vega have been used, and data have been corrected for galactic reddening and aperture effects. $M_{v}$ has been calculated using a distance modulus of 32.87.}  
\begin{flushleft}
\begin{tabular}{lllllllllllllll}
\noalign{\smallskip}
\hline
\noalign{\smallskip}
Object & x-coord & y-coord & $m_{814}$ & $\sigma_{814}$ & $m_{555}$ & $\sigma_{555}$ & $m_{439}$ & $\sigma_{439}$ & $m_{336}$ & $\sigma_{336}$ & $M_{v}$  & $(v-i)$ & $(b-v)$ & $(u-b)$ \\
\noalign{\smallskip}
\hline\noalign{\smallskip}
 2  &  233.3  &  297.0  &   20.90  &   0.03   &  19.91   &  0.02   &  21.32  &   0.04  &   20.23   &  0.06   &  -13.0   &  -0.99   &   1.41  &   -1.08  \\ 
 3  &  234.0 &  307.0   &   21.72  &   0.12   &  21.59   &  0.13   &  22.27  &   0.17  &   21.01   &  0.20   &  -11.3   &  -0.13   &   0.68  &   -1.26 \\  
 4  &  240.1 &  314.6   &   21.00  &   0.04   &  20.19   &  0.03   &  21.26  &   0.06  &   19.89   &  0.05   &  -12.7   &  -0.81   &   1.07  &   -1.37 \\  
 5  &  257.0 &  299.0   &   21.46  &   0.05   &  22.29   &  0.08   &  22.85  &   0.14  &   22.44   &  0.38   &  -10.6   &   0.83   &   0.56  &   -0.40 \\  
 6  &  257.0 &  272.5   &   21.40  &   0.10   &  20.42   &  0.03   &  21.52  &   0.08  &   20.25   &  0.07   &  -12.4   &  -0.98   &   1.10  &   -1.27 \\  
 7  &  260.0 &  256.0   &   20.99  &   0.07   &  21.59   &  0.09   &  21.78  &   0.10  &   20.50   &  0.10   &  -11.3   &   0.59   &   0.20  &   -1.28 \\  
 8  &  265.4 &  260.6   &   21.78  &   0.12   &  21.56   &  0.07   &  21.47  &   0.07  &   19.97   &  0.05   &  -11.3   &  -0.22   &  -0.09  &   -1.51 \\  
 9  &  261.0 &  272.0   &   21.63  &   0.07   &  21.08   &  0.06   &  22.35  &   0.14  &   20.72   &  0.09   &  -11.8   &  -0.55   &   1.26  &   -1.62 \\  
10  &  273.0 &  262.0   &   22.84  &   0.17   &  23.11   &  0.21   &  23.20  &   0.20  &   23.43   &  0.75   &  ~-9.8   &   0.28   &   0.08  &    0.23 \\  
11  &  278.0 &  272.5   &   22.53  &   0.27   &  23.03   &  0.30   &  22.78  &   0.17  &   21.59   &  0.24   &  ~-9.8   &   0.51   &  -0.26  &   -1.19 \\  
12  &  289.5 &  274.0   &   21.53  &   0.05   &  22.11   &  0.08   &  22.24  &   0.11  &   21.14   &  0.15   &  -10.8   &   0.58   &   0.13  &   -1.10 \\  
13  &  296.6 &  261.0   &   20.78  &   0.04   &  20.98   &  0.04   &  20.89  &   0.06  &   19.61   &  0.06   &  -11.9   &   0.20   &  -0.09  &   -1.28 \\  
14  &  298.0 &  266.5   &   21.63  &   0.14   &  21.58   &  0.11   &  21.67  &   0.14  &   19.68   &  0.07   &  -11.3   &  -0.05   &   0.08  &   -1.98 \\  
15  &  298.9 &  270.0   &   21.14  &   0.06   &  21.17   &  0.05   &  21.11  &   0.09  &   19.74   &  0.06   &  -11.7   &   0.03   &  -0.06  &   -1.38 \\  
16  &  305.3 &  275.0   &   21.49  &   0.07   &  22.03   &  0.10   &  22.18  &   0.13  &   21.24   &  0.14   &  -10.8   &   0.54   &   0.15  &   -0.93 \\ 
17  &  308.2 &  269.9   &   22.03  &   0.11   &  22.92   &  0.22   &  22.65  &   0.16  &   22.37   &  0.57   &  -10.0   &   0.89   &  -0.27  &   -0.27 \\  
18  &  325.0 &  273.0   &   20.06  &   0.02   &  19.77   &  0.02   &  19.66  &   0.02  &   18.16   &  0.01   &  -13.1   &  -0.29   &  -0.11  &   -1.49 \\  
19  &  277.0 &  281.0   &   20.53  &   0.03   &  19.70   &  0.04   &  20.84  &   0.05  &   19.51   &  0.06   &  -13.2   &  -0.83   &   1.14  &   -1.33 \\  
20  &  281.3 &  282.0   &   20.77  &   0.06   &  19.93   &  0.05   &  21.00  &   0.08  &   19.65   &  0.07   &  -12.9   &  -0.84   &   1.06  &   -1.35 \\  
21  &  310.3 &  289.8   &   21.00  &   0.04   &  20.06   &  0.03   &  21.10  &   0.05  &   19.73   &  0.04   &  -12.8   &  -0.94   &   1.04  &   -1.36 \\  
22  &  270.7 &  307.1   &   22.23  &   0.08   &  21.12   &  0.09   &  21.95  &   0.08  &   20.54   &  0.09   &  -11.7   &  -1.10   &   0.82  &   -1.40 \\  
23  &  285.0 &  323.5   &   17.32  &   0.01   &  17.41   &  0.01   &  17.31  &   0.01  &   15.89   &  0.01   &  -15.5   &   0.09   &  -0.10  &   -1.42 \\  
24  &  294.1 &  329.0   &   21.40  &   0.18   &  21.63   &  0.17   &  21.87  &   0.21  &   20.66   &  0.20   &  -11.2   &   0.23   &   0.24  &   -1.21 \\  
25  &  295.0 &  336.0   &   20.77  &   0.08   &  21.41   &  0.20   &  21.59  &   0.15  &   20.40   &  0.13   &  -11.5   &   0.65   &   0.18  &   -1.19 \\  
26  &  290.3 &  336.8   &   21.36  &   0.11   &  22.23   &  0.40   &  22.45  &   0.24  &   21.88   &  0.58   &  -10.6   &   0.87   &   0.22  &   -0.57 \\  
27  &  305.0 &  335.0   &   19.40  &   0.03   &  19.49   &  0.05   &  19.44  &   0.03  &   18.16   &  0.04   &  -13.4   &   0.09   &  -0.05  &   -1.28 \\  
28  &  303.3 &  342.0   &   19.23  &   0.02   &  19.44   &  0.02   &  19.53  &   0.02  &   18.35   &  0.03   &  -13.4   &   0.21   &   0.09  &   -1.18 \\  
29  &  315.8 &  342.0   &   19.75  &   0.02   &  20.20   &  0.03   &  20.14  &   0.03  &   18.96   &  0.03   &  -12.7   &   0.45   &  -0.05  &   -1.18 \\  
30  &  313.0 &  254.0   &   21.95  &   0.07   &  22.28   &  0.12   &  22.18  &   0.10  &   20.89   &  0.11   &  -10.6   &   0.33   &  -0.10  &   -1.29 \\  
31  &  310.5 &  256.6   &   22.20  &   0.10   &  22.12   &  0.12   &  22.06  &   0.09  &   20.71   &  0.10   &  -10.8   &  -0.08   &  -0.05  &   -1.35 \\  
32  &  282.0 &  357.0   &   20.96  &   0.07   &  21.22   &  0.08   &  21.12  &   0.06  &   19.72   &  0.05   &  -11.6   &   0.26   &  -0.10  &   -1.40 \\  
34  &  300.1 &  354.9   &   21.32  &   0.15   &  21.05   &  0.10   &  21.69  &   0.15  &   20.54   &  0.20   &  -11.8   &  -0.27   &   0.64  &   -1.15 \\  
35  &  300.8 &  359.6   &   21.04  &   0.10   &  20.69   &  0.07   &  20.97  &   0.07  &   19.69   &  0.08   &  -12.2   &  -0.35   &   0.28  &   -1.28 \\  
36  &  303.0 &  362.9   &   20.94  &   0.08   &  20.79   &  0.06   &  20.95  &   0.05  &   19.86   &  0.07   &  -12.1   &  -0.15   &   0.16  &   -1.09 \\  
37  &  295.0 &  363.6   &   20.65  &   0.05   &  21.12   &  0.07   &  21.08  &   0.06  &   20.16   &  0.10   &  -11.8   &   0.46   &  -0.04  &   -0.92 \\  
38  &  285.7 &  342.0   &   22.32  &   0.28   &  22.30   &  0.34   &  22.58  &   0.28  &   21.53   &  0.39   &  -10.6   &  -0.02   &   0.27  &   -1.04 \\  
39  &  281.1 &  343.7   &   21.61  &   0.13   &  21.78   &  0.20   &  22.04  &   0.21  &   20.69   &  0.14   &  -11.1   &   0.17   &   0.26  &   -1.35 \\  
40  &  277.0 &  342.4   &   21.78  &   0.14   &  21.15   &  0.09   &  21.93  &   0.15  &   20.41   &  0.13   &  -11.7   &  -0.63   &   0.78  &   -1.51 \\  
41  &  271.0 &  339.6   &   21.94  &   0.17   &  21.63   &  0.09   &  22.14  &   0.17  &   20.53   &  0.12   &  -11.2   &  -0.31   &   0.51  &   -1.61 \\  
42  &  311.0 &  319.7   &   22.36  &   0.20   &  22.26   &  0.16   &  23.50  &   0.43  &   22.43   &  0.52   &  -10.6   &  -0.10   &   1.24  &   -1.07 \\  
43  &  283.7 &  379.0   &   20.68  &   0.03   &  21.38   &  0.05   &  21.49  &   0.05  &   20.54   &  0.07   &  -11.5   &   0.71   &   0.11  &   -0.95 \\  
44  &  309.0 &  374.0   &   19.37  &   0.01   &  20.03   &  0.02   &  20.29  &   0.03  &   19.27   &  0.03   &  -12.8   &   0.66   &   0.26  &   -1.02 \\  
45  &  322.9 &  383.0   &   21.07  &   0.08   &  21.06   &  0.08   &  21.37  &   0.09  &   19.73   &  0.08   &  -11.8   &  -0.01   &   0.32  &   -1.64 \\  
46  &  324.2 &  388.0   &   20.38  &   0.04   &  19.60   &  0.02   &  20.29  &   0.03  &   18.92   &  0.03   &  -13.3   &  -0.78   &   0.69  &   -1.37 \\  
47  &  342.3 &  385.3   &   20.81  &   0.04   &  20.52   &  0.04   &  21.08  &   0.05  &   19.88   &  0.06   &  -12.4   &  -0.29   &   0.57  &   -1.20 \\  
50  &  326.0 &  403.3   &   20.51  &   0.08   &  19.69   &  0.07   &  20.49  &   0.09  &   19.05   &  0.10   &  -13.2   &  -0.82   &   0.80  &   -1.44 \\  
51  &  325.6 &  400.3   &   20.63  &   0.06   &  19.80   &  0.04   &  20.63  &   0.05  &   19.07   &  0.05   &  -13.1   &  -0.83   &   0.83  &   -1.56 \\  
52  &  335.0 &  401.6   &   20.97  &   0.08   &  20.58   &  0.05   &  21.90  &   0.15  &   20.65   &  0.13   &  -12.3   &  -0.39   &   1.32  &   -1.24 \\
53  &  343.3 &  427.5   &   19.46  &   0.02   &  18.38   &  0.01   &  19.48  &   0.02  &   18.06   &  0.01   &  -14.5   &  -1.08   &   1.10  &   -1.41 \\  
54  &  337.0 &  434.0   &   21.35  &   0.06   &  21.57   &  0.08   &  22.67  &   0.15  &   21.64   &  0.21   &  -11.3   &   0.22   &   1.10  &   -1.03 \\  
55  &  350.0 &  423.4   &   21.50  &   0.11   &  20.73   &  0.07   &  21.58  &   0.09  &   20.40   &  0.10   &  -12.1   &  -0.76   &   0.85  &   -1.18 \\  
57  &  326.0 &  414.0   &   22.17  &   0.12   &  21.54   &  0.11   &  22.99  &   0.27  &   21.88   &  0.30   &  -11.3   &  -0.63   &   1.45  &   -1.10 \\  
58  &  310.1 &  411.7   &   21.95  &   0.10   &  22.61   &  0.19   &  22.46  &   0.11  &   21.19   &  0.14   &  -10.3   &   0.65   &  -0.15  &   -1.27 \\  
59  &  325.0 &  429.0   &   22.28  &   0.07   &  23.14   &  0.16   &  23.94  &   0.44  &   23.47   &  1.02   &  ~-9.7   &   0.86   &   0.80  &   -0.47 \\  
80  &  335.7 &  359.0   &   22.45  &   0.10   &  23.68   &  0.27   &  23.74  &   0.36  &   23.90   &  1.50   &  ~-9.2   &   1.23   &   0.06  &    0.16 \\  
83  &  330.8 &  319.0   &   23.22  &   0.16   &  23.81   &  0.39   &  24.63  &   0.77  &   ~~-     &  ~~-    &  ~-9.1   &   0.58   &   0.82  &    ~~-  \\ 
85  &  285.0 &  292.0   &   21.86  &   0.11   &  20.67   &  0.05   &  21.55  &   0.08  &   20.06   &  0.06   &  -12.2   &  -1.19   &   0.88  &   -1.49 \\  
86  &  248.4 &  325.0   &   22.95  &   0.13   &  21.88   &  0.06   &  22.65  &   0.13  &   21.18   &  0.13   &  -11.0   &  -1.07   &   0.77  &   -1.47 \\  
87  &  254.3 &  333.7   &   22.59  &   0.12   &  22.24   &  0.09   &  23.06  &   0.18  &   21.99   &  0.31   &  -10.6   &  -0.35   &   0.81  &   -1.07 \\  
88  &  281.0 &  253.3   &   22.87  &   0.10   &  22.67   &  0.11   &  22.19  &   0.07  &   20.89   &  0.12   &  -10.2   &  -0.20   &  -0.48  &   -1.30 \\ 
\hline
\noalign{\smallskip} 
\end{tabular}
\end{flushleft}
\end{table*}

\subsection{Results of photometry}

Figure 4 displays the the photometric uncertainties as a function of apparent magnitude for the four used filters in the outer and inner sample. The photometric uncertainty is the one calculated by {\tt phot} and includes only photon statistics; i.e. the other error discussed above are not included.
In Fig. 5 we present a colour-magnitude diagram of the objects in the inner and outer sample. The sources, on average, become redder with decreasing luminosity, although there is considerable scatter in the relation. 

In Fig. 6 a $(b-i)$ vs. $(b-v)$ colour-colour diagram is presented for the inner sample and those objects in the outer sample with a combined photometric uncertainty
\footnote{The combined photometric uncertainty in two filters $x$ and $y$ with photometric uncertainties $\sigma_x$ and $\sigma_y$ is: $\sigma_{xy} = \sqrt{{\sigma_x}^2+{\sigma_y}^2}$}     
, $\sigma_{b-v}$, less than 0.2 magnitudes in $(b-v)$. This diagram shows the effect of [OIII]$_{\lambda5007}$ emission line contamination. At low ages the $(b-v)$ colour will be quite red due to the contribution from  [OIII]$_{\lambda5007}$ to the $v$ magnitude. Most objects follow the prediction from the photometric evolution models (Sect. 5) but some objects have excess flux in the $v$ band as compared to the model. 

\begin{table}
\caption[]{Median photometric properties of the outer, inner and total sample. Total, is the combined inner and outer sample.}
\begin{flushleft}
\begin{tabular}{llllll}
\noalign{\smallskip}
\hline\noalign{\smallskip}
sample & $M_v$ & $(u-b)$  &  $(b-v)$ & $(b-i)$ & $(v-i)$ \\
\noalign{\smallskip}\hline
\noalign{\smallskip}
inner  & -11.5 & -1.21 & 0.33 & 0.28 &-0.04   \\
outer  & -9.5 & -0.68 & 0.29 & 0.85 & ~0.82 \\
total  & -10.6 &-1.13 & 0.30 & 0.49 & ~0.52\\
\noalign{\smallskip}
\hline
\end{tabular}
\end{flushleft}
\end{table}

\begin{figure}
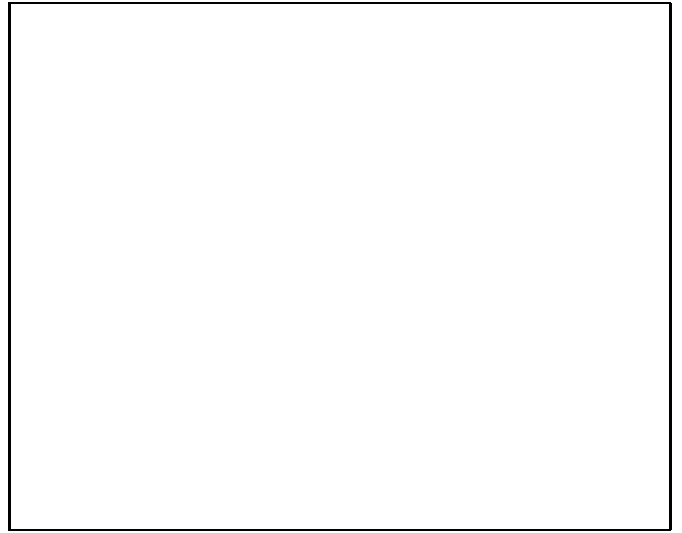

\picplace{7cm}
\caption[]{ 
Colour-colour diagram of the objects in the  $(b-i)$ vs. $(b-v)$ space. Outer sample objects with photometric errors $\sigma_{b-i}<0.1$ are shown as large filled circles and outer objects with $0.1 \le \sigma_{b-i}<0.2$ as smaller filled circles. Inner sample objects are shown as open circles, also with their size depending on the photometric accuracy. A photometric evolution model, with a Salpeter IMF and 10 \% solar metallicity (see Sect. 5) is plotted as a dotted line, with pentagrams at ages 1, 10, 100, 1000 and 11000 Myr. The 1 Myr location is close to $(b-i)=0~$ and $(b-v)=0.7~$ and the evolutionary track ends close to $(b-i)=1.8~$ and $(b-v)=0.8~$ at 11 Gyr. One can clearly see two distinct features: Most objects, especially the outer ones, are found close to the track predicted by the photometric model. Many inner objects, however, are found to be situated above the predicted curve, suggesting  that [OIII]$_{\lambda5007}$ contamination in the F555W filter is important. The arrow in the lower right of the plot is the effect of internal reddening by $E(B-V)=0.1~$, i.e. since no internal reddening has been applied to the data here, the data points should be moved in the opposite direction of the arrow, if there was internal extinction. }
\end{figure}

 Fig. 7 presents a $(u-b)$ vs. $(v-i)$  diagram for all objects where these colours could be measured. In Fig 8. we show the luminosity function and the $(u-b)$, $(b-v)$ and $(v-i)$ distributions for the outer and inner sample respectively. Table 3  presents the median values of $M_v$, $(u-b)$, $(b-v)$ and $(v-i)$ for the two sub-samples. The luminosity and colour distributions will be discussed in Sect. 6.

\begin{figure}
\picplace{7cm}
\caption[]{
Colour-colour diagram of the objects in the $(v-i)$ vs. $(u-b)$ space. All objects which were detected in all four pass-bands are included. Outer sample objects with photometric errors $\sigma_{b-i}<0.1$ are shown as large filled circles and outer objects with $0.1 \le \sigma_{b-i}<0.2$ as smaller filled circles. Inner sample objects are shown as open circles, also with their size depending on the photometric accuracy. The reddening vector in the lower right corner corresponds to  $E(B-V)=0.1~$, c.f. caption Fig. 6. }
\end{figure}

\begin{figure*}
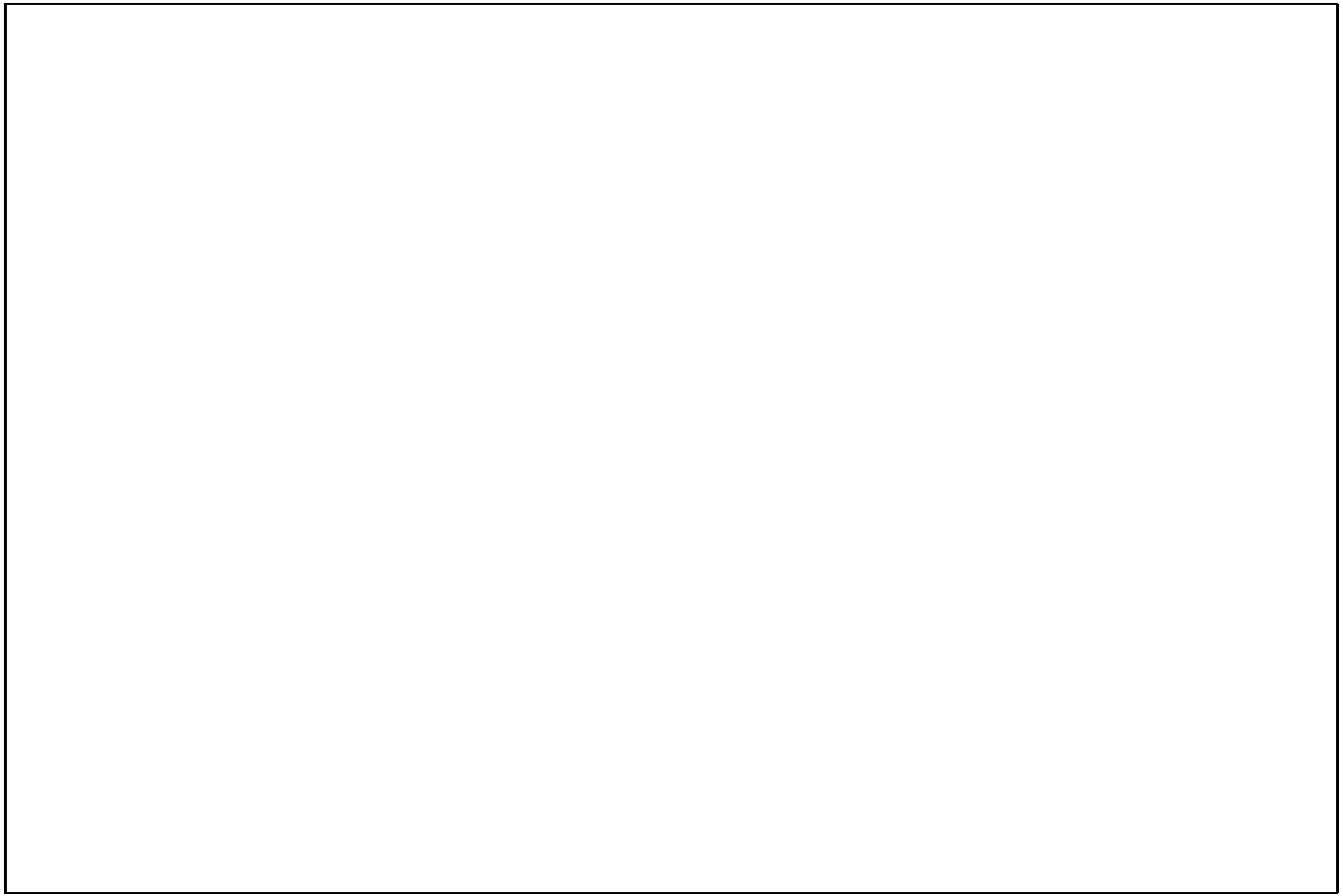

\picplace{12cm}
\caption[]{
Luminosity functions and colour distributions: The upper panel shows from the left the $M_v$, $(u-b)$, $(b-v)$ and $(v-i)$ distributions for  the outer sample. The bin-sizes are, 0.5, 0.125, 0.2 and 0.125 magnitudes respectively. Only objects with measured magnitudes are included. Therefore e.g. the ($u-b$) diagram may be biased to the blue since some redder objects detected in F814W and F555W but not in F439W or F336W, are not included. The lower panel shows the same for the inner sample.  }
\end{figure*}

\section{Are the globular cluster candidates associated with ESO~338-IG04?}

There is an apparent concentration of faint point-like objects near ESO~338-IG04.
This was realised already from studying ground based images, and was one motive for studying this galaxy with HST.
The apparent over density of faint objects was the first reason why these were suspected to be associated with ESO~338-IG04. 
 As will be shown in this section, our new HST observations confirm the physical association of these faint objects with ESO~338-IG04. There is a clear concentration and over density of faint point-like objects around the galaxy (Figs. 2 and 9). The distribution  shows a spatial flattening which agrees closely with  the isophotal shape of the galaxy at radii greater than 150 pixels (1.2 kpc), see Fig. 9. This suggests a coeval formation of the galaxy and the population of outer objects and that the system is dynamically relaxed.

\begin{figure}
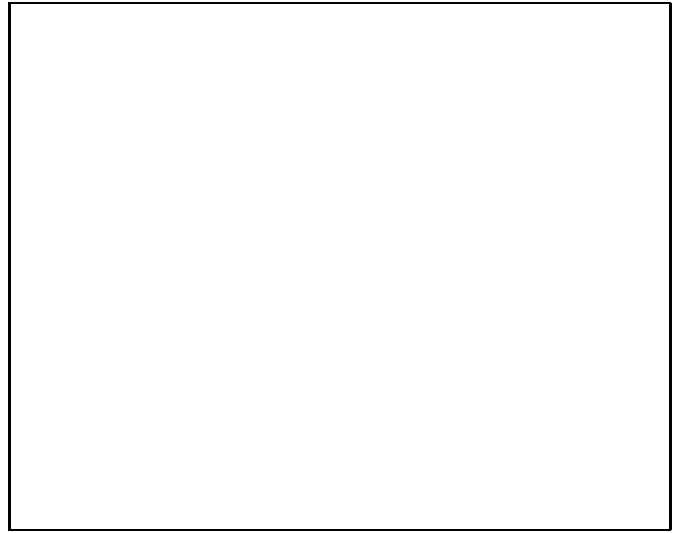

\picplace{7cm}
\caption[]{Histogram showing the number of objects as a function of distance from the centre of the galaxy. All objects in the outer sample with $m_{814} < 24 $~ mag are included. The dashed line represents what would be expected for a constant surface density distribution, i.e. for objects not associated with the galaxy. The dotted line shows the smoothed radial luminosity distribution in the $V$ band (obtained from ground based observations, Bergvall and \"Ostlin 1998), in units of total flux per radius bin (non logarithmized). Thereby the luminosity distribution is directly comparable to the number distribution of globular cluster candidates as a function of distance. For small radii the radial distribution of clusters does not follow the luminosity distribution. The reason is that the outer sample is incomplete (since we are in a transition region between the outer and inner sample) and the galaxy has excess surface brightness due to the active starburst. At radii greater than 150 pixels the radial distribution of the objects closely follows the luminosity distribution of the galaxy. This strongly supports that the objects belong to the galaxy, and that the galaxy and the objects have a common physical origin. The pixel scale of the PC is 0.0455 $\arcsec$/pixel, which at the assumed distance of 37.5 Mpc corresponds to 8 pc/pixel.
}
\end{figure}

\subsection{Expected number of Galactic foreground stars}

Utilising the galaxy model by Bahcall and Soneira (1981) we find that the predicted number of foreground stars brighter than $m_V=24$ magnitudes is  approximately one within our $\sim 30\arcsec\times 30\arcsec $ PC field.
Thus the expected contamination of foreground stars is very small. However the Bahcall and Soneira model does not include the Galactic halo. Therefore we also examined the adjacent wide field camera (WFC) WFC2 and WFC3 fields of the WFPC2 (WFC4 contains a larger fraction of the faint extensions of the target galaxy and was not included) and counted all stellar objects brighter than the $24{\rm th}$ magnitude in $m_{555}$ and $m_{814}$. In all, we found  67 objects with $m_{814}<24$, and 28 objects with $m_{555}<24$. The PC area coverage is 1/10 of that of two WFC chips, excluding the vignetted areas, so we would then suspect 6.7 and 2.8 stars in our PC field in F814W and F555W respectively from this comparison. This is however an overestimate since on the WFC chips some hits/hot pixels are left which will be hard to distinguish from faint stars due to the large pixel scale. Also the bright part of the target galaxy covers more than 1/10th of the area on the PC, and any faint foreground star would completely drown in the light from the bright starburst. Furthermore, it is possible that some sources in the WFC fields near the border to the PC field are actually objects associated with ESO~338-IG04. Taking this into account we  expect to find less than 5 and 2 foreground stars brighter than the $24{\rm th}$ magnitude in F814W and F555W respectively. Also, since the number of foreground stars per magnitude interval increases when going to fainter magnitudes, they will primarily appear at the faint red end of our distribution, while most of our discussion will consider  brighter objects.  We conclude from these two estimates, that not more than a few objects are expected to be galactic stars, and in that case primarily some of the faint red ones. One obvious foreground star (which in fact is an apparent double) is located approximately five arcseconds to the west of the centre of ESO~338-IG04 (Fig 2).

\subsection{Photometric distances}

Another way to estimate the probability that some of the objects could be foreground stars is to calculate their photometric distances. By assuming them to be main sequence stars with spectral class given by their $(v-i)$ colour, we can determine their absolute magnitudes and calculate their distance modulii. For this purpose we used the $(V-I)$ versus $M_V$ relation by Jarrett (1992). For all outer objects for which we measured both the $m_{814}$ and $m_{555}$ magnitude, the photometric distances were calculated. All but 3 (out of 70) objects have photometric distances larger than 13 kpc, excluding the possibility of them being Milky Way stars. Three objects have lower photometric distances: object no. 34 (9 kpc), no. 39 (6 kpc) and no. 45 (10 kpc); these could possibly be halo stars. If we would  assume the objects to be giants, the photometric 
distances would increase by approximately a factor of ten, so the only possibility of them being foreground objects would be that they were all white dwarfs. Considering that white dwarfs constitute only a small fraction ($\sim 5 \%$, Gliese, 1982) of the local stellar luminosity function, we conclude that it is very unlikely that even one could contaminate our sample. A few sources with $m_{814} > 24$ are not visible in the V band images; these could be foreground stars or background galaxies. In the inner sample all objects have photometric distances in excess of 20 kpc, except no. 44 for which the photometric distance is 14 kpc. N.B. that a low photometric distance does not at all imply that the object really is a foreground star, only that $(v-i)$ and $m_{555}$ is consistent with a such.

\subsection{Expected number of background galaxies}

A few very faint diffuse objects have been found in the images, which probably are background galaxies. To asses the possibility that bright nuclei of distant galaxies have entered our sample of globular cluster candidates we have used the recent result from Hubble Deep Field by Williams et al. (1996). The expected number of galaxies brighter than 23.75 magnitudes is $\approx 4$ within the PC aperture (also considering the obscuration of the central starburst) for the F814W images and less in the F555W images. Considering that only galaxies with bright point-like nuclei are likely to cause confusion (also their total magnitude must then be brighter then 23.5 lowering the number even more), we conclude that our results are most probably not affected by background galaxies.  This is also supported by counting the galaxies in the adjacent WFC chips.

We see one obvious, fairly bright, galaxy NW of the main body, with the interesting feature that it  is only visible in the F814W images. The discussion of this object will be devoted to a separate investigation. Other possible  galaxies are  object number 14 and 39 in Table 1.

We could of course be extremely unfortunate, so that ESO~338-IG04 was projected onto a very distant (with a redshift above 1) cluster of galaxies lurking just below the detection threshold. In that case bright elliptical galaxies or bulges could enter the faint part of our sample. We have measured the spatial extent of all objects in our outer and inner sample and found that they are generally far too small to be galaxies. If part of the faint end was contaminated from a background cluster there should be a grouping of the faint objects, since these are generally redder, unless ESO~338-IG04 was centred on the cluster. No grouping in colour can be found in the distribution, which indicates that a background cluster of galaxies do not contaminate our sample, as expected.    

There are a few objects which appear so red that they are only visible in the F814W image. These are the best candidates for being background galaxies, since the colours are redder than expected for globular cluster. Alternatively, these could be foreground stars.

\subsection{Spatial extent of globular cluster candidates}

In this subsection we deal with the sizes of the GC candidates. One should bear in mind that even with the PC pixel size, the PSF of the telescope is under-sampled. Moreover, we are trying to estimate sizes typically smaller than the pixel size. Therefore these estimates are necessarily uncertain  and the fitted size of one single object should not be over-interpreted.

\subsubsection{Cross correlation method}

In order to see if the sources are extended or not we have used the following method.
The images were cross-correlated with model PSFs and a PSF convolved modified power-law 2D cluster profiles (Lugger et al., 1995). The model profiles are good approximations to a $c=2$ King model, at a radius $<$ 14 core radii ($r_c$).
The method which is applicable to low signal-to-noise data is similar to the
the cross-correlation method for surface photometry by Phillips and Davies (1991). The normalised cross-correlation is maximised when the source and the model have the
same light profile.  
Model PSFs were generated with TINYTIM, including "jitter"  data for the images. The flux centra of the convolved model images were aligned to the pixels with the highest fluxes in the source images. 
  In no case is the cross-correlation maximised for the bare PSF. A typical $r_c$ is 0.5
pixels for the F555W data and 0.3 pixels for the F814W data, which corresponds to 4 and 2.5 pc respectively at the distance of ESO~338-IG04. The derived $r_c$ for an object in F555W and F814W is the same within 0.2 pixels, in all cases. The method is sensitive to variations in the background light distribution. The reason for F555W data giving larger sizes could be due to   emission line contamination and larger background noise in this filter.
 No reliable results were
obtained for sources overlaid on the main body of ESO~338-IG04, i.e. the inner sample. The method will generally overestimate sizes, since noise will make peaks appear less sharp on average and equal weight is given to all pixels within the fitting window.

A number of tests were performed to check the
reliability of this method. For example we ran the method on stellar images in the field.   
As there are no suitable stars in the PC chip this test was done on the WFC fields. For stellar images the 
cross-correlations are always maximised for $r_c \sim$ 0.0-0.1 pixels.
 In Fig. 10 a histogram of the estimated core radii of outer clusters in the F814W filter is shown. The distribution peaks at 2.5 pc. 
The majority of Galactic GC's have $r_c < 2$~pc with a peak in the distribution slightly below 1 pc (Harris 1995). Thus our GC candidates have larger radii than the average galactic GC, but fall within the observed distribution. 

Translating a core  radius (half surface brightness) to effective  radius (half-light), depends sensitively on the adopted $c$ value of the King-model used. For small $c$ the difference between the effective and core radius is small, while it is large for large values of $c$. For galactic globular clusters $c$ ranges between 0.7 and 2.5 with 1.5 being a typical value. We have used $c=2$, since it has an analytic representation at small radii (Lugger et al. 1995) convenient for model fitting. Since Milky Way GCs have a range in $c$ values, with $c=2$ being a relatively high value, we cannot directly convert our distribution in core radii
to a distribution in effective radii. Galactic GCs with ~$ c \approx 1.5$~  have ~$r_h/r_c \approx 2$~ (Trager et al. 1993, Harris 1996). Some authors (e.g. Whitmore et al. 1993) have used gaussian profiles for which the core and effective radius are equal (half width half maximum of the gaussian profile). Using a gaussian is equivalent to using a very small value of $c$, and will lead to an overestimate of the core radius if the true $c$ value would be larger than one.

We also experimented using gaussian profiles convolved with the PSF. This yielded core radii (which here equals effective radii) approximately a factor of two larger than when using $c=2$~ King profiles, with a median value of 6 pc. Comparing this to the distribution of effective radii for galactic GCs, which peaks at at $\sim$2.5 pc and has a median value of 3 pc (Djorgovski 1993,  Harris 1996), we find again that our clusters are approximately a factor of two larger. This also indicates that the choice of of ~$c$~ is not crucial. 

\subsubsection{Growth curve method}

To derive the spatial size of cluster candidates, many authors have used the magnitude difference in apertures of varying size,  i.e. something close to a growth curve (e.g. Whitmore et al. 1993, Schweizer et al. 1996). This gives a measure of the width of the light profile, i.e. the spatial extent. To check for consistency we applied also this method on the candidate GCs. For all objects we measured $\Delta m_{0.8-3.0}$, the magnitude difference between two small apertures with radii 0.8 and 3 pixels respectively; the same estimate was used by Schweizer et al. (1996). We measured this quantity for the convolved King profiles, the sole PSF and the cluster candidates. In addition we also measured $\Delta m_{0.8-3.0}$ for PSFs convolved with gaussian profiles. Since this method uses information from a smaller area than the cross-correlation method it is less sensitive to background fluctuations. Therefore the inner sample objects posed no problem and their spatial extent could be assessed.

For the  GC candidates we obtained values of  $\Delta m_{0.8-3.0}$ ranging from $1.0$ to $2.5$, with a median value of $\Delta m_{0.8-3.0}=1.5$ for the outer sample and $\Delta m_{0.8-3.0}=1.7$ for the inner sample. Thus there is a weak tendency for the inner sample objects to be larger. Comparing the distribution of $\Delta m_{0.8-3.0}$ values, the inner sample has a less peaked distribution. This may indicate that a few sources in the inner sample  are  not GCs but more extended associations, but it may also reflect the larger difficulties in measuring the inner sample objects. Projection effects may further cause objects to appear more extended than they are. The median $\Delta m_{0.8-3.0}$  values correspond to core radii of 0.25  and 0.4 PC pixels respectively. This is in perfect agreement with the measurements from the cross correlation method for the outer sample. At a distance 37.5 Mpc the median  core radii estimates  correspond to 2.0 and 3.2 pc  and the quartile ranges in $r_{\rm c}$ (which contain 50 percent of the objects) are 1.3-1.8 and 1.4-5.6 pc for the outer and inner sample respectively. These values are fully consistent with GCs.

When comparing the median  $\Delta m_{0.8-3.0}$ values to PSFs convolved with gaussian profiles we derive median effective radii (which for a gaussian profile equals the core radius) of 0.56 and 0.80 PC pixels for the outer and inner sample respectively. This correspond to  4.5 and 6.4 pc at the distance 37.5 Mpc. This is very close to the median effective radius of $5$ pc found by Schweizer et al. (1996) for GC candidates in the merger remnant NGC~3921. Our $\Delta m_{0.8-3.0}$ vs. effective radii  relation for the F814W filter is almost identical to the one derived by Schweizer et al. (1996) for the F555W filter. Convolving our Tiny Tim PSFs for F555W with gaussians yield a  $\Delta m_{0.8-3.0}$ vs. effective radii relation which is offset with 0.1 magnitudes in  $\Delta m_{0.8-3.0}$ at a given effective radius, as compared to Schweizer et al. (1996). It is possible that Schweizer et al. (1996) actually used PSFs for F814W rather than F555W as was claimed, or that there are substantial differences in the Tiny Tim models and PSFs measured in stellar fields. The difference between our and Schweizer et al.'s relations are however unimportant since we use the F814W filter.  

It is difficult to assess whether the broader distribution of $r_{\rm c}$ for the inner sample is a consequence of larger intrinsic spatial extent or primarily reflects the general photometric difficulties in the inner sample region. For many sources in the inner sample the distance (centre to centre) to the nearest neighbour is only on the order of five pixels, and in some cases less. This means that the outer used aperture in the  $\Delta m_{0.8-3.0}$ may contain a considerable contribution from neighbouring objects. The effect of this will an overestimate of the spatial extent of the inner objects with close neighbours, i.e. almost all inner objects. A conservative estimate can be obtained by removing the most extended sources in the outer sample until the median value and width (as measured by the standard deviation) of the inner sample agrees with that of the outer sample. To achieve this 27 percent of the inner sources must be removed. Hence even assuming that crowding does not affect the distribution of core radii, only a minor fraction of the inner objects would have to be discarded as GC candidates. To further investigate this we measured also  $\Delta m_{0.8-1.5}$ for the convolved King profiles and the outer and inner objects. Now the inner sample objects have a more peaked distribution of fitted core radii and the difference between the median $r_{\rm c}$ of the inner and outer sample decrease. This indicates that crowding may be the main cause for the larger appearance of the inner objects. An aperture with radius 1.5 will for some sources still be contaminated with light from a neighbouring source. In fact both samples get smaller median core radii: 1.75  and 2.7 pc for the  outer and inner sample respectively. That also the outer sample radii decrease indicate that King $c=2$~ profiles are not perfect representations for the sources in general.

In the outer sample two sources, object number 32 and 64 (see Table 1), appear as sharp as the PSF, i.e. unresolved and therefore suspected stars. A few  more sources: number 39, 45 and  58 are only marginally resolved, and objects 13, 18 and  19  are resolved but the uncertainty in the $\Delta m_{0.8-3.0}$ measurement overlaps with the PSF.  However objects number 18, 39 and 58 are clearly resolved using the GC method. In the inner sample, object no. 24 appears unresolved. Two objects in the outer sample (no. 63 and 68) and 9 in the inner sample(no. 4, 9, 10, 21, 22, 35, 39, 51 and 55) have  $\Delta m_{0.8-3.0} > 2$, corresponding to $r_{\rm c} > 6 $ pc. 

The cross-correlation and $\Delta m_{0.8-3.0}$ methods are found to yield fully consistent results. The estimated radii of the clusters fall within the distribution observed for Galactic GCs. The inner sample objects are on average more extended, reflecting the greater difficulty assessing their sizes and perhaps that a small part of the inner sample objects are associations rather than young GCs. However, it should be stressed once more, that we are working with objects which have most of thier light concentrated within one PC pixel. Thus all estimates of their spatial extent will be  uncertain, no matter what method is used.

The peak in our distribution of core radii occurs at 2.5 pc, approximately a factor 2 larger than the Milky Way distribution. Could the larger size be a simple selection effect because  the  clusters we observe generally are more massive and thus on average larger than galactic ones? Probably not,  since Galactic GCs show no clear correlation between  radius and luminosity(van den Bergh et al. 1991, Harris 1996). 

\begin{figure}
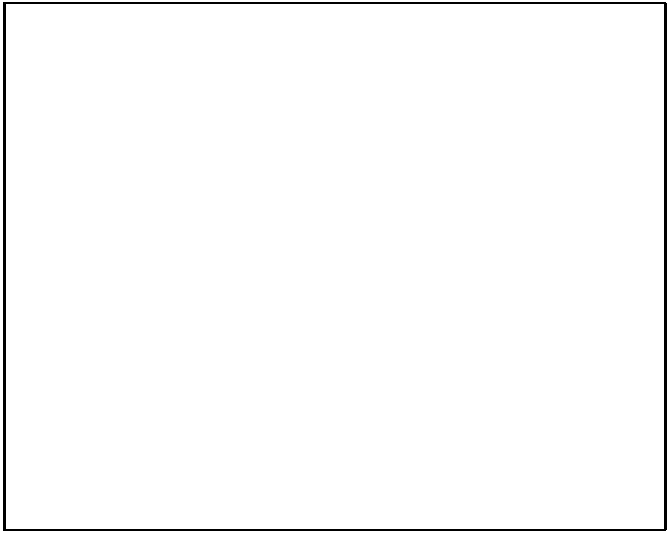

\picplace{7.0cm}
\caption[]{Histogram of the estimated core radii of outer clusters with good signal to noise in the F814W filter, using the cross correlation method. the distribution peaks at 2.5 pc. The bin-size in this plot is 0.2 pixels, or 1.65 pc. }
\end{figure}

\subsection{Could some objects be super giants in ESO~338-IG04?}

The distribution of absolute magnitudes of the GC candidates is shown in Fig. 8.
All  objects in Table 1 and 2 are brighter than $M_v =-7$. Comparable luminosities can be reached by super giant (SG) stars, i.e. massive stars in their late evolutionary stages. It is thus important to assess if our sample of GC candidates can be contaminated by luminous SGs in   ESO~338-IG04. 

To compare our objects with properties of individual stars, we have used the same evolutionary tracks as in the photometric evolution model, described in sect. 5.1. We used the transmission profiles of the used filters convolved with  the CCD response curve to calculate the magnitudes of the model stars in the VEGAMAG system. Thereby, we are able to study the location of stars in the $M_v - (v-i) - (u-b)$ space and compare with the location of our GC candidates.
We find that luminosities of corresponding to $M_v =-7$ or greater can only be reached or stars with mass greater than approximately 15 ${\cal M_{\odot}}$. 
However, to match also the observed $(v-i)$ colour, only stars more massive than  20 ${\cal M_{\odot}}$ can cause confusion. Such  stars have life times shorter than about 10 Myr. 

The current star formation activity is strongly concentrated to the centre of ESO~338-IG04. If we assume that a star has a typical velocity of $\sim 10 $km/s, it would  need more than 10 Myr to travel the distance from the centre to the region of the outer sample. Thus it is unlikely that the outer sample will be contaminated with SGs that has emigrated from the centre. However some star formation appears to be going on, at a slower pace, over an extended region outside the central star burst, why the presence of SGs is still possible. As we showed in the last section, the majority of the objects are spatially extended, which makes it unlikely that they are individual stars, unless they have associated nebulosity or reside in clusters; in the latter case they are of course not single stars at the present resolution.
For the inner sample region, we expect several SGs to exist, but we will only be able to distinguish them as individual stars if they are well isolated. Since star formation tends to be a collective phenomenon, producing not only one star at an instant, we expect star formation to occur in associations and clusters. 

By studying the location of the model stars and our observed objects in the $M_v - (v-i) - (u-b)$ space, we identify two areas where our objects overlap with the location of stars. The first one is for blue super giant (BSG) stars with $M_v < -9.5, -0.25 < (v-i) < 0.0$ and $-1.6 < (u-b) < -0.9$. Only stars more massive than 60 ${\cal M_{\odot}}$ will ever have such properties, and in total four of the objects are found in this region: no. 91 in the outer sample and no. 31, 42 and 88 in the inner sample. 

The other region where SGs and our objects partly overlap is for $M_v < -7$ and $  (v-i) > 0.3$. In this region, not all sources have measured $(u-b)$ colours so for these we have less constraints on their nature. For those objects with measured $(u-b)$ colours only one, no. 2 and to some extent no. 53 in the outer sample, have properties consistent with a red/yellow super giant (RSG) star. For those without $(u-b)$ measured, no firm conclusions can be drawn, but only stars more massive than 20 ${\cal M_{\odot}}$ will ever reach comparable luminosity and $(v-i)$ colour. Speaking against their identification as RSGs rather than GCs is their location in the outer sample, and that the majority appear spatially resolved.

Luminous RSGs  origin from stars in the mass range 20 to 40 ${\cal M_{\odot}}$, and the duration of the RSG phase is on the order of 10 000 years. From the observed $H\alpha$ luminosity and a Salpeter IMF we expect a star formation rate of approximately one ${\cal M_{\odot}}$ per year in the area covered by the PC field of view.  For a Salpeter IMF,  a fraction of approximately 0.0005 of the newly formed stars  will have a mass in the range required for it to become a RSG. Thus we expect on the order of a dozen luminous RSGs to be present in ESO~338-IG04 at each instant. The number of BSG luminous enough to enter the region in the $(v-i) - (u-b)$
plane where the GC candidates reside can in the same way be estimated to be less than five. Thus we expect several SG stars to exist  in ESO~338-IG04, but they are more likely part of the clusters than observable as individual stars, and in any case, their number should be ten times smaller than the number of observed GC candidates. Object no. 2 and 91 in the outer sample and no. 31, 42 and 88 in the inner sample, are the best candidates for being individual SGs based on their colour and luminosity; however all appear to be spatially extended, and only no. 2 in the outer sample is beter fit by a single SG than with the photometric GC model described in Sect. 5.
Almost a third of the objects in the outer sample lack measured $(u-b)$ colour and are from the measured $(v-i)$ and $M_v$ consistent with the predictions for RSGs. A few of these, no. 13, 19, 32, 45 and 64,  are in addition unresolved or only marginally resolved (Sect. 4.4.2), making the RSG interpretation possible for these.

\subsection{Conclusions on association}

We conclude from the analysis above that the overwhelming majority of objects seen around ESO~338-IG04 in the PC image must be associated with the galaxy. In general, we have shown that foreground star contamination is improbable and  that the photometric distances are inconsistent with foreground stars. 
Moreover, the vast majority of object have colours and absolute magnitudes that excludes the possibility that they
could be super giant stars in the target galaxy. Although SGs should be present, we expect them to be part of clusters and associations rather than isolated objects. Furthermore, isolated giant stars should not appear spatially resolved. The remaining alternative is then that these are actually star clusters in ESO~338-IG04. Since many of them appear to be quite old (see next section) and their light distribution is concentrated it is most likely that they actually are globular clusters.

Can we single out any probable interlopers? In the next section we will see that the photometric evolution models predict $(v-i) \la 1.1$ for old GCs. Objects which within the 1 sigma uncertainty are redder than this are therefore unlikely to be GCs, unless they have very high metallicity or internal reddening. This is the case for objects 39, 45, 13, 19 and 63 in the outer sample. In addition objects 14 and 64 are too red for $(v-i)$ to be measured. Of these, all but no. 14 and 63, have $\Delta m_{0.8-3.0}$ values whose 1 sigma uncertainties include the PSF. In view of this it is likely that they are foreground stars, while no. 14 and 63 may be background galaxies. However all but no. 39 and 45 have photometric distances in excess of 10 kpc which cast some doubt on the foreground star interpretation, unless they reside way out in the halo. An alternative is that some of these could be individual RSGs in ESO~338-IG04 after all. We should also add object no. 32 in the outer sample to the interloper candidates, since it appears unresolved and in addition, very red. However, not all of these objects have colours that are fully consitent with main sequence stars.  Thus the nature of these eight objects remain unclear, although they are not likely to be GCs in ESO~338-IG04. In number they constitute only five percent of the total number of objects in the outer and inner sample, and affect in no way the conclusions in this paper.   

\section{Ages of globular cluster candidates}

\subsection{Spectral evolutionary synthesis models}

To estimate the ages of the cluster candidates we used a spectral evolutionary synthesis code developed by Bergvall. This is based
on evolutionary tracks of stars of metallicity in the range 5\% (mainly) to 10\% solar. Tracks
in the mass range 0.08-120 ${\cal M_{\odot}}$were obtained from
Schaller et al. (1992). Pre-main sequence tracks were obtained from 
VandenBerg (1985, 1986), Iben (1965) and Grossman and Graboske (1971). 
Horizontal-branch and AGB tracks up to the onset of 
thermal pulsation were obtained from Castellani et al. (1991).  Equivalent widths of
the Balmer lines and HeI lines in absorption for stars with
temperatures above 30000K were obtained from the non-LTE stellar 
atmosphere models of Auer and Mihalas (1972). Line data for cooler stars 
were obtained from
different sources in the literature. The evolutionary tracks were 
combined with synthetic stellar spectra
obtained from Kurucz (1992). At the highest effective temperatures Kurucz' 
data were supplemented with data from Howarth and Lynas-Gray (1989; 
computed from the Kurucz model) covering a wider range in log g. Data for 
nebular line emission were obtained 
from Ferlands models (1993) and the continuous emission was calculated 
from Brown and Mathews (1970) and Sibille et al. (1974). The metallicity of 
the gas was assumed to be 10\% solar.  With this set-up we were able to make 
predictions of galaxy spectra as a function of time and initial mass function (IMF).

These spectra have been integrated for the HST F336W, F439W, F555W and F814W filters using the 
transmission curves available at the STSCI web-site. As for the photometry, the zero-points
for Vega were used. To check the accuracy a spectrum of Vega was integrated the same way
and yielded zero colours.
Since we derived the model colours using the transmission profiles of the filters, convolved with the CCD sensitivity, and the photometric zero-points for Vega (HST data handbook)  together with a spectrum of Vega, our estimates do not rely on any colour terms. Furthermore, since we use the full filter profiles the results are insensitive to red-leaks of the HST filters. 

Both the contribution from line- and continuous emission are included in the model and thus, contamination from gaseous emission should not be a problem. Unfortunately, this will not be the case since the emission is, at least partly, spatially decoupled from the ionising source, and  found in the form of filaments several tens of pc away. This may be a problem for the inner sample. Moreover, contrary to e.g. H$\alpha$, the predicted equivalent width of [OIII]$_{\lambda5007}$ is sensitive to metallicity and to unknown physical parameters like the filling factor of the gas in the nebula. 

The models produce predicted fluxes for each time step which are fit to the observed fluxes
of the cluster candidates using a least squares method.
 The models also, for a given IMF,
produce mass to luminosity ratios ($M/L_v$) for the stellar population, including the mass contained in stellar remnants. These values can be used to estimate the mass of each cluster candidate by simply multiplying the $M/L_v$ value at the best fitting age with the observed absolute V-band flux, expressed in solar units.  

\subsection{Model parameters}
 
We have assumed a "top hat" star formation history in the models. The star formation rate is constant for $\tau_{\rm SF}= 10$ Myr and zero thereafter. Since we are modelling regions of limited spatial extent such
a short time-scale is probably realistic. Our results are not sensitive to the value of $\tau_{\rm SF}$ as long as it is shorter than the age of the object. In regions where star formation is still going on, the choice of $\tau_{\rm SF}$ will affect the modelled ages.
We used the model predictions at the time steps  10, 20, 30, 40, 50, 60, 70, 80, 90, 100, 200, 300, 400, 500, 600, 800, 1000, 1500, 2500, 3500, 5000, 7000, 9000 and 10800 Myr to fit to the observed data. 

\begin{figure}
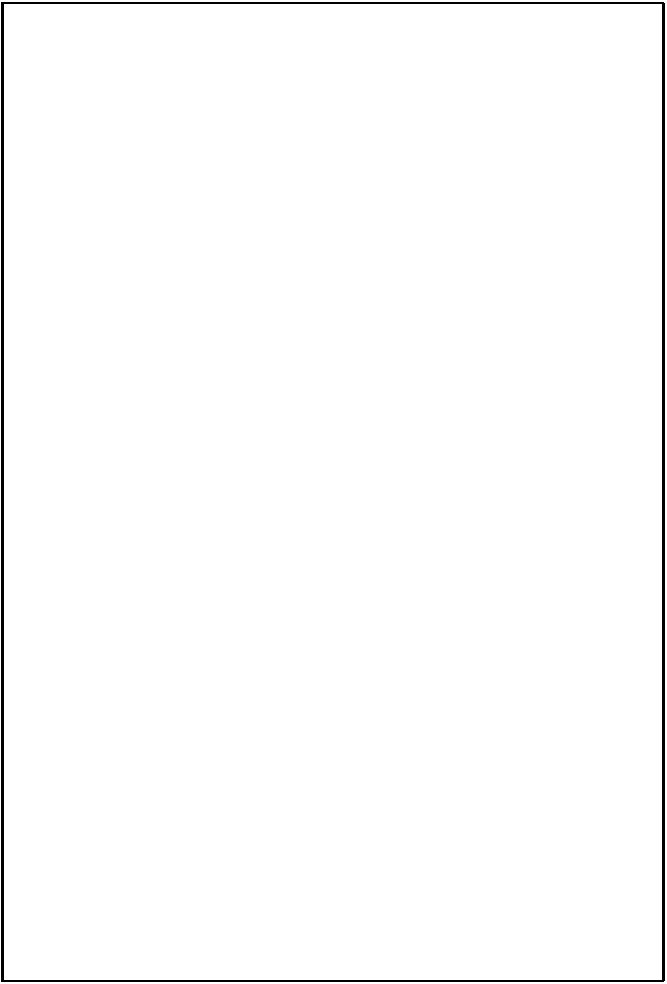
 
\picplace{13.0cm}
\caption[]{Photometric evolution of the models used. The x-axis is the logarithm of the age in years. The top panel shows the evolution of $M_v$ for the four different IMFs. The solid line represents the prediction for the Salpeter IMF ($\alpha=2.35$), the dashed line for $\alpha=1.35$, the dotted line for $\alpha=3.35$ and finally, the dash-dot line shows the Miller-Scalo IMF. On the rightmost axis the mass to luminosity ratio is shown for the same models. The second, third and fourth panel from the top shows the evolution of $(u-b)$, $(b-v)$ and $(v-i)$, respectively. All colours correspond to the HST photometric system with zero-points for Vega.  }
\end{figure}

Since the number of observables in the data set is too small to unamibigously discriminate between different IMFs we will focus our discussion on results based on the classical Salpeter IMF  (Salpeter 1955) and the more realistic Miller-Scalo IMF (Miller-Scalo 1982).
The Salpeter IMF is a single power-law with  $\alpha=2.35$, the power-law index\footnote{The IMF is defined as $ dN/dM \propto M^{-\alpha} $, where $M$ denotes the stellar mass and $dN$ the number of elements in the interval $[M,M+dM]$.}. The Miller-Scalo  IMF is characterised by three different power-laws at different mass intervals:  $\alpha=3.30$  for $[10{\cal M_{\odot}},120{\cal M_{\odot}}]$,  $\alpha=2.50$ $[1{\cal M_{\odot}},10{\cal M_{\odot}}]$ and $\alpha=1.40$ for $[0.1{\cal M_{\odot}},1{\cal M_{\odot}}]$.
 To illustrate how dependent our results are to the chosen IMF, we have also modelled the objects using a flatter ($\alpha=1.35$) and a steeper ($\alpha=3.35$) mass function. We will sometimes refer to these as the $flat$ and $steep$ IMFs respectively. In Fig. 11 we show the photometric evolution of a model globular cluster from 1 Myr to 11 Gyr for the four different IMFs. As can be seen, the major difference occurs at low ages. In particular the steep IMF gives considerable redder $(v-i)$ colours, due to the relatively larger influence of pre-main sequence evolution (assuming "naked" pre-main sequence stars). Another major difference is the difference in the mass to luminosity ratio, $M/L_v$. As we will see, only the Miller-Scalo IMF gives a  $M/L_v$ value that agrees with those observed for old galactic GCs. 

\subsection{Metallicity}

The assumed metallicity is 5-10\% ([Fe/H]$\approx$-1.2) and 10\% of the solar value for the stars and gas, respectively. The metallicity of the gas was chosen to be close to the observed nebular oxygen abundance of $12\%$ solar in ESO~338-IG04. The metallicity of the gas is only important at low ages when the ionizing field from young stars give rise to recombination radiation from the gas. The choice of metallicity for the stellar component is less obvious, since it may vary with the age of the stars. The young, newly formed clusters will have metallicity close to the nebular value, while possible old objects may have a considerably lower metallicity. It is well known that metallicity has a major impact on the colours of a stellar population, making them redder with increasing metallicity, and therefore it has to be handled with care. Models using a higher metallicity, e.g. solar, generally gives very different results.

	Observations of elliptical galaxies reveal that several have bimodal
colour distributions, which is interpreted as two populations with similar 
age but different metallicity. In general, the metal rich GC population has
a metallicity close to that of the host galaxy, and a lower peak magnitude
in the luminosity function (Forbes et al. 1997). Durrell et al. (1996) studied
the globular cluster systems of dwarf ellipticals in the Virgo cluster. 
Assuming them to have similar ages as Galactic GCs, they found the mean 
metallicity of the GC systems in the Virgo dwarfs to be [Fe/H]=-1.45,  
for a host galaxy luminosity of $M_v \sim -17$. Subtracting the starburst
in ESO338-IG04, the underlying galaxy has a similar absolute luminosity, 
$M_v \approx -18$.

Thus, there is no reason to assume that the metallicity of any GC in ESO~338-IG04 should be higher than 10 \%. The metallicity of Galactic GCs vary between [Fe/H]=-2.5 and $\approx 0.0$, with a clear peak at [Fe/H]=-1.5, but with no dependence on absolute magnitude (i.e. mass).   Depending on the age of our objects, we expect the metallicity to lie in the range 
[Fe/H]=-2 to -1, and the mean metallicity should be somewhere around  [Fe/H]=-1.5 as in dwarf ellipticals (Durrell et al. 1996) and the Milkyway (Harris 1996). Our used metallicity of [Fe/H]$\approx$-1.2 is thus not far from this. Objects that form very early, during the early collapse phase of the protogalaxy may have a very low ([Fe/H] $\approx-2.5$) metallicity while only slightly older objects may be one order of magnitude more metal rich.  With this uncertainty in the metallicity of the GC candidates we will have corresponding uncertainties in $(v-i)$ of less than ~$\pm 0.1$~ magnitudes at a given age (Couture et al. 1990). The uncertainties in  $(b-v)$ and $(u-b)$ will be comparable. Population synthesis models, e.g. Worthey (1994) predict somewhat less metallicity dependence on the colours at low metallicities. It should be borne in mind, that part of the structure we find in the age distribution of GC candidates, may  be illusive and caused by varying metallicity among the objects. 

\begin{figure}
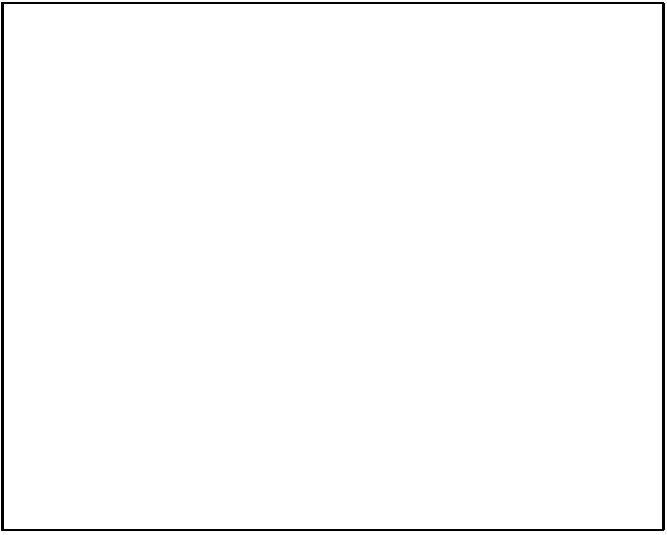

\picplace{7.0cm}
\caption[]{Photometric evolution of the used models, in the  $(v-i)$ versus $(u-b)$ plane. The full drawn line with pentagrams represents the Salpeter IMF model. The pentagrams indicate ages of 1 Myr, 10 Myr, 100 Myr, 1 Gyr and 11 Gyr, with the lowest and highest age in the lower left and upper right corners respectively. The dashed line shows the same for the flat ($\alpha=1.35$) IMF and the dotted line with triangles the steep IMF ($\alpha=3.35$) model. Shown as filled circles with thick error bars, are outer sources with combined photometric errors in F439W and F336W, $\sigma_{u-b} < 0.2$. The length of the error-bars corresponds to the combined one sigma photometric errors in $(v-i)$ and $(u-b)$. Shown as open circles with thin  error-bars are inner sources with $\sigma_{u-b}<0.1$. Here the error-bars are the 1 sigma combined errors multiplied with 1.6 and 1.3 for $(v-i)$ and $(u-b)$ respectively. 
The arrow in the lower right part of the diagram is the reddening vector for an extinction $E(B-V)=0.2$. No internal extinction correction has been applied to the data points in this diagram.
}
\end{figure}

\subsection{Internal extinction}

The colour of any astronomical object will be affected by interstellar extinction. The reddening caused by our galaxy has been corrected for as discussed above. Any additional internal extinction in ESO~338-IG04 will of course affect our results. The internal extinction in the centre of the galaxy is very small;
Bergvall (1985) finds $E(B-V)=0.05$, Iye et al (1987) $E(B-V)\le0.04$ and Calzetti et al. (1995) $E(B-V)=0.02$. It may be a reasonable assumption that this is true also for the cluster candidates, since the metallicity is low and the amount of dust might be small outside the central component of the host galaxy. Some dust will be produced internally in the clusters, due to mass loss from stars in their late evolutionary stages. However, Galactic GCs appears to have very little internal extinction, suggesting that internally produced dust somehow is removed from a cluster (cf e.g Origlia et al. 1997). All the extinction measurements above concerns only the bright starburst which has a high ultraviolet flux. Dust may have been photo-dissociated in this region, while dust outside the main starburst region, i.e. where our outer clusters reside, may have survived. A further uncertainty concerns the extinction law, which appears to be different in starburst galaxies (Calzetti et al., 1994). On the other hand, the outer clusters are not residing in the starburst itself, whence the average Galactic or  LMC extinction laws  probably are more adequate here.  
It is also likely that clusters have different amounts of internal extinction  depending on their age and on the physical conditions during their formation.

One possibility to assess the internal extinction is to vary it and determine where the residual from the fitting procedure for each object is minimised. Since the F336W and F439W data have rather low signal-to-noise for most sources, this may be hazardous and the fitted extinctions may finally rather reflect the uncertainties in the data.
A safer way would be to model the cluster candidates with different internal extinctions for the whole sample to see where the average residual has its minimum. For the Salpeter IMF the minimum average residual for the whole outer sample was obtained for an extinction of $E(B-V)=0.09$. For the flat ($\alpha=1.35$), steep ($\alpha=3.35$) and Miller-Scalo IMFs the corresponding $E(B-V)$ values were 0.08, 0.0 and 0.10 respectively. These number should be viewed critically, since they  may reflect intrinsic systematic errors in the models, but the values are reasonable. Moreover we found that adding or subtracting an extinction of this magnitude had a negligible impact on the overall result.
As a compromise, and due to the lack of information, an internal extinction of $E(B-V)=0.05$ and the average interstellar extinction law (the difference between this and the LMC extinction law is small in the optical region) was assumed for all objects. This value is reasonable in view of galactic GCs.
In Fig. 12 it can be seen that an internal extinction correction of $E(B-V)\sim 0.1$ would bring the data into close agreement with the models, except for the bluest objects. For the bluest sources the extinction has to be small to make a good fit to the models. As long as the assumed internal extinction is small, changing it or the extinction law has no significant impact on the modelled properties, and the resulting age and mass distributions are almost identical.

\subsection{Modelled cluster ages and masses}

To estimate the ages of the globular cluster candidates we used the observed fluxes, as described above, with an internal extinction correction of $E(B-V)=0.05$. 
For each time-step in the model, a least squares fit is made to the observed fluxes.  
The observed fluxes in each filter are weighted with the inverse of their photometric uncertainty. 
This procedure gives a best fitting age for each cluster
candidate, where the residual, $Res$, defined as the weighted mean square deviation between the model and the data, is minimised. 

\begin{eqnarray}
Res = \sqrt{\frac{\sum_{k} \Delta_{k}^{2} w_k}{\sum_{k}w_k}} , \mbox{~where} \nonumber \\
 \Delta_{k}^2= \left( \frac{f_{\rm obs,\it k}-f_{\rm mod,\it k}}{f_{\rm obs,\it k}} \right)^2; 
 \mbox{~here~~} w_k = (\sigma_k)^{-2}.
\end{eqnarray}

The symbols $f_{\rm obs,\it k}$ and $f_{\rm mod,\it k}$ represent the observed and model flux in the band-pass $k$, and $w_k$ is the corresponding weight.

In effect, the F555W and F814W data are generally given higher weights than the F439W and F336W data. Thus, it is not a problem if a particular data point has a low signal-to-noise, since it is given a lower weight in the fitting procedure.
 As the minimum photometric uncertainty used for weighting in each filter, we assumed $\sigma=0.03$ as an estimate of the intrinsic photometric errors with WFPC2 which is of that order of magnitude. For the inner objects where [OIII]$_{\lambda 5007}$  contamination may be a problem, the weights for the F555W data were reduced with a factor of two. 

Figure 12 shows a $(v-i)$ versus $(u-b)$ diagram for objects with good photometry together with the model predictions. (The motivation for only including objects with accurate photometry is for reasons of clarity; in Fig. 7 all sources are included.) The outer sample objects are shown as filled circles with thick error-bars whose length correspond to $\sigma_{v-i}$ and  $\sigma_{u-b}$, the 1-sigma combined errors\footnote{$\sigma_{v-i}=\sqrt{\sigma_{555}^2+\sigma_{814}^2}$, $\sigma_{u-b}=\sqrt{\sigma_{336}^2+\sigma_{439}^2}$ } in $(v-i)$ and $(u-b)$. The inner sample objects are shown as open circles with thin error-bars. The length of the error-bars for the inner sample corresponds to 
$1.5\sigma_{v-i}$ and  $1.3\sigma_{u-b}$. The inner sample formal errors have been scaled to make a comparison with outer sample objects more fair (Sect 3.2). It can be directly seen that objects span a wide range of ages, with inner objects generally being younger. The agreement between the models and the observed objects is very good. As discussed above, adding an extinction correction of $E(B-V)=0.1$ would make the agreement even better, except for the youngest objects. Although there are a few objects with blue $(u-b)$ and red $(v-i)$ colours, close to the prediction of the steep IMF, most objects appear to be better fitted by the other IMFs. If there was a bimodality in the IMF, this would be very interesting, but it is not possible to claim this from our data, considering the various uncertainties. For the objects with the bluest $(v-i)$ colours, it appears that only the flat IMF model can reproduce them. This does not necessarily mean that the flat IMF is to prefer, since we are now dealing with ages shorter than $\tau_{\rm SF}$, and with sources which all have strong emission line contamination, and further the distance to the prediction at the lowest age in the Salpeter IMF is only a few sigma.    

We have seen that the colour of an object will be determined  mainly by its age, but that the situation is complicated by varying extinction and metallicity, which can have a non neglible impact on the colours.  Below, we will begin  with interpreting the colours as caused by age, and in the next subsection discuss how varying extinction and metallicity might modify the results.

\subsubsection{The outer sample}

For the standard Salpeter IMF the resulting age distribution for the outer objects is shown in Fig. 14. We did not include objects 14 and 64 in the modelling since they were not detected in the F555W, F439W or F336W data (14 was actually detected in F336W, which suggests that it  may be a distant galaxy).
We can identify what appears to be at least four different epochs in the cluster formation history. There is a group of young clusters with ages from $\le$10 to 100 Myr, an intermediate age  200-700 Myr group, an older 1-7 Gyr group with a peak at 2.5 and 5 Gyr and  finally an ~$\sim$11 Gyr old  group.  Thus there are indications of populations of young as well as old globular clusters present in this galaxy. Varying the IMF does not affect the general pattern in the age distribution. Although the ages of the objects shift slightly ant the relative strength of the peaks change slightly, they remain at the same locations. All IMFs produce the same median age of 2.5 Gyr.

The histogram of modelled masses are shown in Fig. 14 for the Salpeter IMF. It can be seen that the clusters are very massive, with a peak in the distribution around $10^6$ ${\cal M_{\odot}}$. Even the objects with the lowest modelled mass, $\ga 10^4$ ${\cal M_{\odot}}$,  are massive enough to be globular clusters. However, the modelled mass of a GC is quite sensitive to the assumed IMF due to the varying $M/L_v$ value (Fig. 11). The median mass of the clusters is $6.7\times10^5 {\cal M_{\odot}}$ for the Salpeter IMF, $3.7\times10^5 {\cal M_{\odot}}$ for the Miller-Scalo IMF, $1.8\times10^6 {\cal M_{\odot}}$ for the steep IMF and $4.2\times10^6 {\cal M_{\odot}}$ for the flat IMF. At an age of 11 Gyr, typical for old Galactic GCs according to recent estimates (e.g. Chaboyer et al. 1998), the different IMFs predict the following $M/L_v$ values: 5.7 for the Salpeter, 3.5 for the
Miller-Scalo, 11 for the steep, and 44 for the flat IMF; all expressed in solar units. The $M/L_v$ values for well studied old Galactic GCs range between 1.8 and 3.6 (Meylan and Pryor 1993). Thus only the Miller-Scalo IMF can reproduce the observed  $M/L_v$ values. The flat and steep IMFs are both clearly ruled out, while the Salpeter IMF is at the limit. 

As can be seen in Fig. 15 there is a correlation between the mass and age such that old clusters on average are more massive.  Since the luminosity of for a given mass decrease with time, i.e. the $M/L_v$ value increases, we do not expect to find any old low mass clusters. Plotted in Fig. 15 is the photometric mass of a cluster with $m_{814}=24$ at a function of age. This is our expected low mass cut-off and agrees well with the observed distribution. Considering that, including the objects falling below the detection threshold, we expect the number of old clusters to be larger than the number of young ones (Table 4). Therefore we expect to find more massive clusters among the older ones. This also means that the apparent situation in Fig. 14, that the total number of clusters is dominated by young objects, most likely is not real (cf Sect. 6.1 and 6.6).

In Fig 14 we also show the luminosity function of all objects in the outer sample, transformed to an age of 9 Gyr using the  Salpeter model, comparable to ages of galactic GCs (Chaboyer et al. 1998). The luminous part of the distribution have a shape not too far from gaussian,  with a peak at $M_v=-8.5$. The shape is similar to the luminosity function of galactic globular clusters but the peak luminosity is $\sim$ 1 magnitude brighter for our objects, due to incompleteness. The estimated absolute magnitude at an age of 9 Gyr is also sensitive to the IMF as seen in Table 4.

\subsubsection{The inner sample}

For the inner objects, it can be seen from their colours, in Figs. 6,7,11  and 12, that all but a few have ages less than 100 Myr. 
In Fig. 14 we  show the resulting age and mass distribution for the inner sample, and also  $M_v$ transformed to an age of 9 Gyr. Masses range from $\sim 2 \times 10^4$ ${\cal M_{\odot}}$ to $\sim 5 \times 10^6$ ${\cal M_{\odot}}$ assuming a Salpeter IMF. The transformed luminosity function does not have the nice shape of the outer sample. The inner sample is incomplete and in addition many sources may be double or multiple clusters, and further photometric errors cannot be neglected. This may explain the irregular shape of the luminosity function at 9 Gyr. Further this transformation only includes photometric evolution, not dynamic. 
Approximately 40 percent of the inner sample objects are fitted with an age $\le 10$ Myr. This subset may be contaminated by associations but most objects are expected to be young GCs since they have small radii and large masses.  The older objects in the inner sample, are probably outer sample objects projected on the centre of the galaxy.

\subsubsection{The total sample}

The division between an outer and inner sample was made on practical grounds, not physical. In studying the statistical properties of the sources it is more interesting to combine the two sub-samples into a $~total~$ sample. Table 4 presents median modelled properties for the whole sample as a function of age of the clusters, for the Salpeter and Miller-Scalo IMFs. Except for the youngest objects, where the two IMFs give almost equal $M/L_v$ values, the Salpeter IMF produce about twice as massive clusters.  We noted in the previous section that only the Miller-Scalo IMF gives 
$M/L_v$ values in accordance with observed GCs. Therefore the flat and steep IMFs are not included, since they give unrealistic $M/L_v$ values for old clusters. Table 4 also presents $<M_{v,9{\rm Gyr}}>$, the median of $M_v$ transformed to an age of 9 Gyr including only photometric evolution. In general the values predicted by the two used IMFs are within a few tenths of a magnitude, with the largest difference again at low age. The quantity $n_{<-8.5}$ is the number of sources with $ M_{v,9{\rm Gyr}} < -8.5$, which is of relevance for assessing the richness of the GC population; and $n$ is the total number of objects in the age interval in question. If there is a possible contamination of at most five foreground/background sources, $n$ should be lowered by this number but not   $n_{<-8.5}$ since the
possible interlopers are all faint. The median mass in each age interval is consistent with GCs. The youngest objects have a lower median mass reflecting the age dependent mass detection limit, see Fig 15.

\subsection{Uncertainties}

In the following section we will critically examine the resulting age distribution obtained in Sect. 5.5. We discuss how reliable our results are considering  photometric, metallicity, extinction and model uncertainties. 

\subsubsection{Photometric errors}

To estimate uncertainties we made the objects bluer and redder with the observed one sigma uncertainties. This had a minor effect on the shape of the age histogram. When adding the errors in the blue and red direction the median age was shifted down to 1500 Myr and up to 3500 Myr respectively. Thus the age pattern in Fig. 14 is very robust to changes in the IMF and to internal errors.  Not included in the modelling, nor in Table 1, are objects with $m_{814} \ga 24$, many of which are only detected in the F814W filter. These may partly be foreground stars, but some of them may be old low luminosity globular clusters. 

The age distribution of the clusters in the outer and inner sample seen in Fig. 14 reveals that objects are found in most age bins. This could mean that clusters have formed at a more or less steady rate since  more than 10 Gyr back in time, and that at some occasions, e.g. at the present, the cluster formation rate have been heavily enhanced. However, it is likely  that the true age distribution is different from the modelled one due to observational uncertainties, and uncertainties in the models. Even if the true distribution is sharply peaked at different ages, the modelled distribution is likely to be smoother, since the corresponding peaks in the colour distributions would be smeared by observational errors. One specific suspicion was that the 11 Gyr peak was not real, but created by 2.5 and 5 Gyr objects which due to observational errors in the red direction piled up at the reddest model age, i.e. at 11 Gyr.  

To test this possibility we simulated the age distribution in the following way: We assumed an input age distribution with two to five sharp peaks, and the observed distribution of $m_{814}$. The colours of the objects of a specific age are given by the model. Errors are now added stochastically according to the observed $m-\sigma$ relation (Fig. 4). These simulated objects are modelled with the photometric evolution model to check if the age distribution of the observed objects can be reproduced. In this way we can determine which peaks are significant. To reproduce the data we need to have peaks in the age distribution at 10 Myr, 30 Myr, $\sim$100 Myr, $\sim$600 Myr, 2.5 Gyr, 5 Gyr and $\sim$11 Gyr. The peak at 100 Myr is only marginally significant. The 2.5 and 5 Gyr peaks could possibly be replaced by a more extended GC formation epoch lasting a few Gyr, but not with a single peak of intermediate age. Out of the seven suspected foreground/background objects (Sect. 4.5), five were best fitted with an age of 11 Gyr. The total number of objects with this fitted age is 14. Thus up to one third of these could be actually foreground/background objects rather than old GCs in ESO~338-IG04.
The 11 Gyr peak could only partly be reproduced by assuming a 5 Gyr peak broadened by observational errors, even when allowing for one third of them to be foreground/background objects. Thus even if the number of very old GCs are less than the model-fitting indicates, objects with this age are required to explain the observed colour distribution. Thus, in addition to the present active GC formation event, we are left with three prominent GC formation epochs in the past: 0.5, 2.5-5 and $\le$11 Gyr ago. However, the 11 Gyr peak may in reality be less peaked and extending to higher ages.

\subsubsection{Effects of metallicity and extinction uncertainties}

Above we found that the observed distribution of colours could be understood by a wide range of cluster ages with several  peaks in the age distribution. The observed range in $(v-i)$ and $(u-b)$ is about two magnitudes, which is far more than expected from variations in the metallicity and extinction alone. Thus we need this wide range of ages to explain the data. However, we cannot rule out that some smaller variations in the colour distribution are caused by varying extinction or metallicity. It is difficult to disentangle effects of metallicity and extinction, since their impact in a colour-colour diagram like Fig. 12 is very similar.

Could, for example, the apparent age structure seen among objects older than 1 Gyr  (Fig. 14) be caused by varying extinction in objects of a uniform age of, say, $\sim 2$~ Gyr? To reproduce the colour distribution we then need to add an additional extinction of more than 0.3 magnitudes in $E(B-V)$ for those objects that first were fitted with an age of 11 Gyr. The average extinction needed for these objects would be $E(B-V) = 0.6$, and $E(B-V) = 0.2$ for the 5Gyr objects . This is far more than expected, since there are no signs of any appreciable extinction in this galaxy. Thus, it is unlikely that the range of ages (2 to 11 Gyr) inferrred for the oldest objects is an illusion caused by varying extinction. For the 2.5 vs the 5 Gyrs peak, this cannot be ruled out, but the required extinction is still higher than expected. 

We saw in section 5.3 that metallicity variations could cause colour variations up to 0.2 magnitudes in $(u-b)$ and $(v-i)$. One could  expect that the metallicity of the objects correlate with their age, with the oldest clusters beeing the most metal poor. This would make them bluer, and create an illusive youth. So if there is a simple relation between age and metallicity among the GC candidates this would rather hide age variations than create illusionary such. Of course the situation may be more complicated than a simple age metallicity relation, e.g. if GC formation was triggred by merging with a gaseous component with a different metallicity than that of the host galaxy. The colour variations among objects with fitted ages above 1 Gyr is still too large to be explained by metallicity variations alone, unless the apparently oldest objects are very metal rich (almost solar), which we showed in Sect. 5.3 was unrealistic. Thus we are left with a real age structure that would require an opposite age-metallicity relationship and in addition appreciable extinction to be avoided. However, it is less certain wether the two connected age peaks at 2.5 and 5 Gyr are real. The difference in the median $(v-i)$ value of the two sub populations is less than 0.1 magnitudes (the difference in $(u-b)$ is larger), which could  be explained by metallicity 
variations of the order of 1 dex. A smaller metallicity difference could be sufficient if the more metal rich clusters in addition had higher internal extinction.    

The conclusions are that fine structures (at the 50 \% level) in the age distributions could equally well be caused by
a metallicity bimodality as by real age differences, but that the overall age structure remains intact, i.e the peaks at $\le 30$ Myr, $\sim$ 600 Myr, 2.5 to 5 Gyr and $\ge 10$ Gyr remain. Due to the unknown metallicity and extinction, the absolute values of the ages are uncertain at  approximately the 50\% level. Note that the main effect of metallicity and extinction is to change the colours, and thus the interpreted age, of an object. The estimated masses and luminosity evolution of the objects will not be importantly affected. 

\subsubsection{Residuals and significance of fitted ages}

Fig. 15 shows a plot of the residual of a fit for the Salpeter IMF versus the weighted  mean error for each object in the outer sample. The weighted mean error, $\sigma_{\rm w}$, is the for each object mean of all the photometric uncertainties weighted with their own inverse, i.e. the signal to noise: 

\begin{equation}
\sigma_{\rm w} = \frac{\sum_{k}  \sigma_k w_k}{\sum_{k} w_k} 
= \frac{\sum_{k}  \sigma_k {1 \over {\sigma_k}^2}}{\sum_{k} {1 \over {\sigma_k}^2}}
.
\end{equation}

\begin{figure}
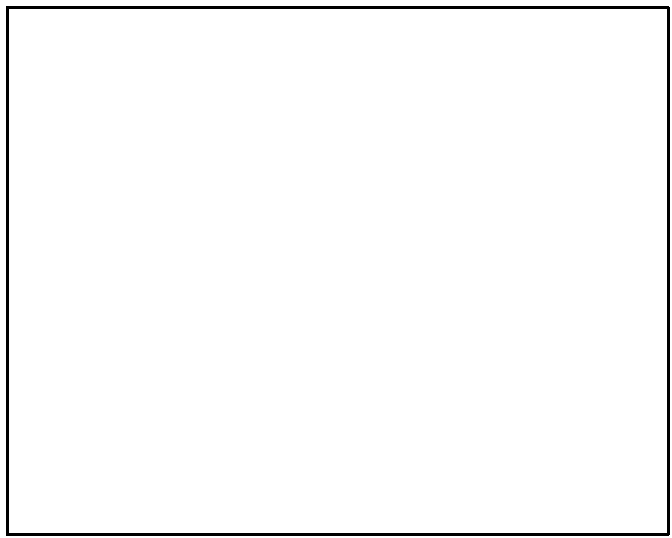

\picplace{7.0cm}
\caption[]{Residuals of fit to the Salpeter ($\alpha=2.35$) model. Left panel: The x-axis shows the  weighted one $\sigma$ mean error for each outer object, and the y-axis shows the residual of the fit. See Eqs. (1) and (2) for the definition of these quantities. For objects below the dashed line the fit is within the 1 sigma photometric uncertainty. Right panel: The ratio of the residual of the fit to the weighted mean  error is shown as a function of the fitted age. Only five objects have fits worse than 2.5 sigmas. Only outer objects have been displayed for reasons of clarity. The inner sample objects show an almost identical distribution of errors and residuals. }
\end{figure}

\noindent
This is the relevant quantity to compare with the residual of the fit, Eq. (1), since it weights the different passbands in the same manner.
On average, we do not expect objects to have smaller residuals than their weighted mean  error. Most objects should however have their residual within a few $\sigma_{\rm w}$. Fig. 15 verifies that this is the case. For  objects with small $\sigma_{\rm w}$, this will underestimate the errors since inherent photometric uncertainties in WFPC2 will enter. It is important to remember that all the given errors in the plots and tables are formal photometric errors, and are therefore lower limits to the true photometric  errors, which may be affected by object centring errors, crowding background subtraction errors, contamination from gaseous emission, inherent photometric uncertainty in the WFPC2 system and possible systematic errors. In particular the former can be a problem for the inner sample. In addition, we have metallicity and extinction and model uncertainties.

Overall, there is no correlation between the fitted age and the residual, except that the lowest ($\le10$Myr) and highest($\ge10$Gyr) age bins have on average slightly higher residuals. This is demonstrated in Fig. 13. This is expected since abnormally blue and red objects will always be fitted with one of these ages although the fit is not particularly good. These objects, where the fit is much worse than expected from the photometric errors are suspect. Perhaps some could after all be distant galactic nuclei or halo stars? In general though, the residuals are well within a few $\sigma_{\rm w}$. For the outer sample and the Salpeter IMF, 65~\% of the objects have fits better than $Res/\sigma_{\rm w}=1$. Only five objects have $Res \ge 2.5 \sigma_{\rm w}$, two which have fitted ages of 10 Myr, one 50 Myr and two 11 Gyr. For the inner sample and the other IMFs the numbers are almost identical.  

Each object has been fitted to one of the time-steps in the model. However, the best fit might have been obtained for an age intermediate between two time-steps. For this reason, objects that apparently coincide with the model prediction e.g. in Fig. 12., may still have a residual larger than the weighted mean error.
We conclude that the residuals from the fit are reasonable as compared to the observational uncertainties, and in general the fits are as good as can be expected.

Although an object will always have a best fitting age, this tell us nothing about the significance of this result. If the signal to noise is sufficiently low, any age would be consistent with the observed properties, and  the fitted age is more or less random. Furthermore, if the residual is only a weak function of the age, these cannot be said to be very significant since several ages are possible within the observational errors. Fortunately, this is generally not the case: most objects have well defined minima in their age versus residual curve. In Fig. 16, the residual for different ages is shown for object 40 in the outer sample, for the four different IMFs used. For this particular object, a Miller-Scalo IMF gives the best fit. The weighted mean  error  for this object is,  $\sigma_{\rm w}=0.04$. It can be seen that for each IMF only a narrow range of ages gives residuals comparable to this.
Most objects have residual-age curves very similar to this with well defined minima.     

\begin{figure*}
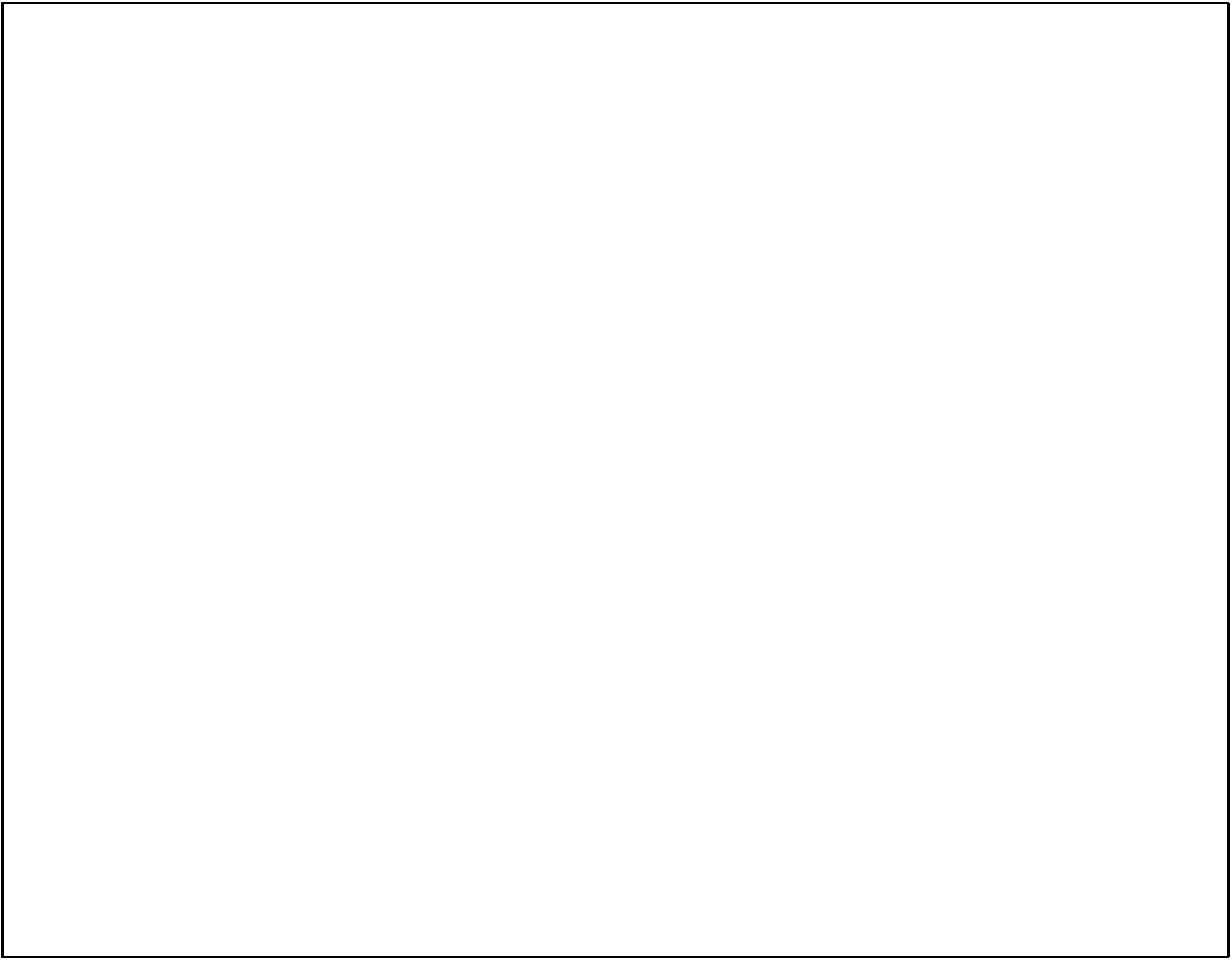

\picplace{14cm}
\caption[]{
Distribution of modelled properties of globular cluster candidates : The top panels shows from the left  the logarithm of the age (in years), the logarithm of the mass (in ${\cal M_{\odot}}$) and $M_{V~9Gyr}$ distributions of the outer sample. The middle and bottom panels show the same for the inner sample and total sample, respectively. NB that the mass detection limit is higly age dependent (cf Fig. 15) favouring detection of young objects. Thus the apparent situation that very young clusters are most numerous is probably illusive.  }
\end{figure*}

\subsubsection{Model uncertainties}

Another important aspect is how accurate the models are. Error sources will come from errors in the input stellar evolution models, stellar atmospheres and from interpolation problems where tracks are sparse or different tracks are patched together. While the latter kind of errors can be controlled, the accuracy of stellar evolution theories are difficult to assess. Charlot (1996) made a comparison of different spectral evolutionary synthesis model and concluded that they may differ appreciably. Furthermore, model spectra generated using the tracks from the Geneva group (the ones used here, Schaller et al. 1992) and the Padova group 
differed in  predicted colours, e.g. $(B-V)$, up to several tenths of a magnitude at certain ages. Therefore, it is very likely that the uncertainties of the model predictions arising from uncertainties in theories of stellar evolution is at least 0.1 magnitudes in colour indices. Our models include AGB stars only to the onset of thermal pulsations. Later stages are very short and not including them will give errors less than 0.05 magnitudes in bolometric flux, and considerable less in the colours (Charlot \& Bruzual 1991).  

\begin{figure}
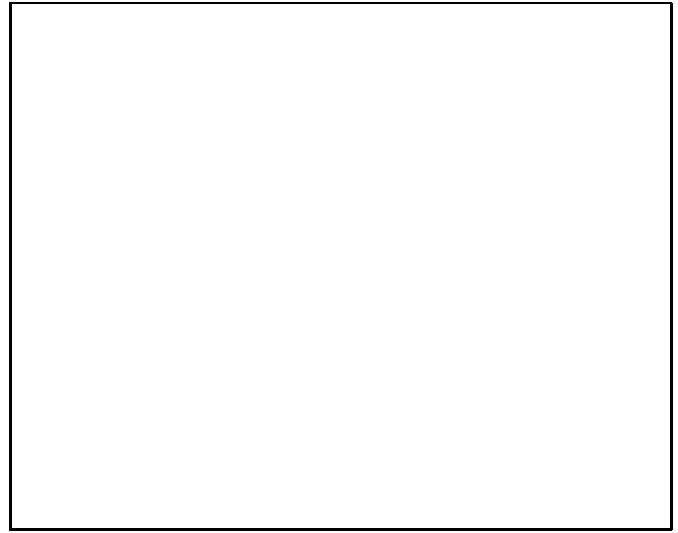

\picplace{7cm}
\caption[]{Relation between modelled age and mass for the outer sample. The dotted line is the photometric mass for an object with given age and $m_{814}=24$, and shows the expected low mass cut-off. The older sources are on average more massive, due to their larger number. However, the inner sample consists of several young massive objects. }
\end{figure}

Further problems are caused by internal extinction, metallicity and uncertainties in the value and uniqueness of the IMF, as discussed above. 
Internal extinction is likely not to be constant, but to vary from cluster to cluster, and the same might be the case for the metallicity. A further uncertainty enters for the youngest objects; here colours are uncertain since the time-scale for star formation, $\tau_{\rm SF}$, is comparable to the modelled ages of the objects. In general $\tau_{\rm SF}$ is not well known. The IMF is a statistical property and for small numbers, e.g. of very massive stars, there will be deviations from the "true" IMF. Since massive stars are very luminous this means that stochastic changes in the mass distributions among the most massive stars can have a noticeable impact on the colours of a young star forming region. Since all our objects are fairly massive, any such fluctuations will be unimportant shortly after the star formation has ceased after 10 Myr. Another effect arises because the model colours and luminosities are smoothed over the width of the time-step. For low ages, rapid evolutionary stages of individual stars will affect a young stellar population. In effect, the ages and masses of clusters with fitted ages 10 Myr and younger have more uncertain ages and masses, than those older than 10 Myr. Some objects will be redder than expected and fitted with a slightly higher age, e.g. 20 Myr, while the too blue objects will still be fit with 10 Myr. At this stage we will not discuss the details of the age distribution within the young objects but are primarily interested in concluding their youth and estimate their masses.

The four IMFs all give comparably good fits on average. The steep IMF gives generally slightly worse fit than the other IMFs, especially for young objects. The three other IMFs give very similar quality of fit, except at low ages where the flat IMF gives better fits. This is due to the presence of some very blue objects as discussed above, and can not be interpreted as clear-cut support for the flat IMF. The strongest support for any individual IMF is the fact that only the Miller-Scalo function gives reasonable mass luminosity ratios at high ages. There appears be a deficiency of blue luminosity at low ages for the Miller-Scalo IMF, indicating that the slope at high masses could be flatter than $\alpha=3.3$. A Miller-Scalo IMF with a slope $\alpha=2.7$ at high masses, would be $\sim 0.3$ magnitudes bluer in $(v-i)$ at ages of a few Myr, and still give a reasonable $M/L$ value at 10 Gyr. We also tried by "hiding" the pre-main sequence stages in circum-stellar dust, with the result that the model $(v-i)$ colours get slightly ($\sim 0.2$ magnitudes) bluer at ages less than 10 Myr. Thus with a few modifications the Miller-Scalo IMF can be made compatible with the bluest observed $(v-i)$ colours.  However, it cannot be excluded the problem with the colours at low ages is related to general uncertainties in the theories of massive star evolution, and to the observational uncertainties.
In view of all these uncertainties, the fits of the objects to the models is as good as can be expected. The main conclusions, that in addition to the present one, there have several previous epochs of massive cluster formation, seem very robust.

\subsection{Conclusions on the modelled ages}

The fit of the observed fluxes of the GC candidates was made using a constant metallicity (5-10 \% of solar) and internal extinction ($E(B-V)=0.05$). The result is that the objects span a wide range of ages, from a few Myr to more than 10 Gyr. In addition there were several peaks i the age distribution (cf Fig. 14), which were located at $~< 30$~ Myr, 100 Myr, 600 Myr, 2.5 Gyr, 5 Gyr and 11 Gyr. The majority of these must be real and cannot be explained as caused by observational or model uncertainties. However, we cannot exclude the existance of metallicity and extinction variations among the objects. Within reasonable limits on the metallicity and extinction variations, these can distort the modelled ages at the 50 \% level. Thus we cannot say with confidence that the 2.5 and 5 Gyr peaks are real as a metallicity bimodality is also possible.  However, we can say that there must be a rich subpopulation of objects with age in the range 2 to 5 Gyr. In addition we can say with confidence that there are objects with age $~< 30$~ Myr, $\sim 10$ Gyr and probably also $\sim 100$ and $\sim 600 ~$Myr. We note that, if metallicity increases with age, the main effect of a metallicity variation will be to hide the real age differences present, rather than create illusionary age differences. 

The modelled masses are mainly sensitive to the adopted IMF. The age and mass distributions are shown in Fig. 14 and Table 4. Since the mass detection limit
is age dependent, young objects are more easily detected. Taking this into account (cf Sect. 6.1 and 6.6) the most numerous subpopulation is that with age 2 to 5 Gyr.

\begin{figure}
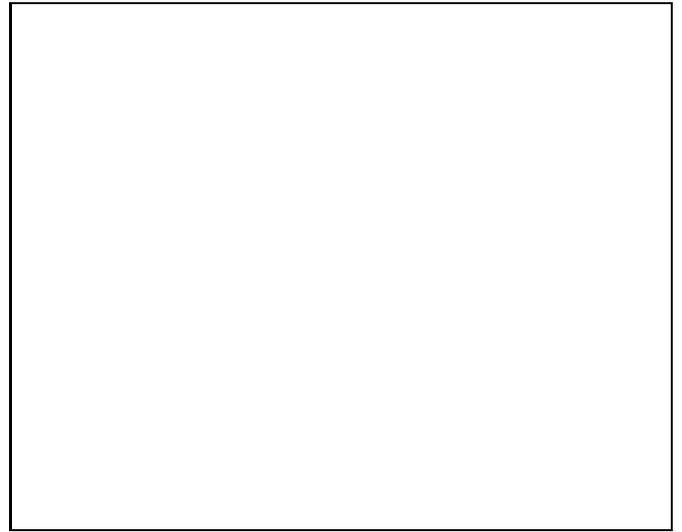

\picplace{7cm}
\caption[]{Relation between age and residual of the fit for object number 40 in the outer sample, for the four different IMFs used. The solid line corresponds to the Salpeter IMF and the dashed, dotted and dash-dot lines to the flat, steep and Miller-Scalo IMFs, respectively. The symbols, pentagrams, circles, triangles and squares correspond to the time-steps used in the modelling. The weighted mean error for this object is $\sigma_{\rm w}=0.04$. Thus only a limited range of ages are consistent with its observed properties. For this particular object the Miller-Scalo IMF gives the best fit. }
\end{figure}

\begin{table}
\caption[]{Modelled properties for the GC candidates in the combined inner and outer sample. $<{\cal M}>$ is the median mass and $<M_{v,9Gyr}>$ the median absolute $v$ magnitude transformed to an age of 9 Gyr. $n_{< -8.5}$ is the number of objects with  $M_{v,9Gyr} \le -8.5$, and $n$ is the total number of objects in each age interval.}
\begin{flushleft}
\begin{tabular}{rlllll}
\noalign{\smallskip}
\hline\noalign{\smallskip}
Age-interval  &&  $<{\cal M}>$   &  $<M_{v,9Gyr}>$ & $n_{< -8.5}$  & $n$  \\
 Myr && $10^5 {\cal M_{\odot}}$ & &   & \\
\noalign{\smallskip}\hline
\noalign{\smallskip}

Salpeter IMF \\
$\le$~50  &&   1.4  & -6.35  &   3   &   57   \\ 
51-300    &&   7.4  & -8.04  &   3   &    8   \\
301-1000  &&   3.6  & -7.31  &   3   &   12   \\
1001-5000 &&   17.0 & -8.95  &   18  &   29   \\ 
$\ge$~5001  &&   9.2  & -8.30  &   6   &   16   \\ 
Whole     &&   4.4  & -7.54  &   33  &  122   \\
 \\
Miller-Scalo IMF \\
$\le$~ 50 &  &   1.6 & -7.07       &  6    &    57  \\
   51-300 &  &   3.5  &  -7.78     &   2  &       8  \\
 301-1000 &  &   2.0  &  -7.24      &   3  &      12  \\
1001-5000 &  &   8.8  &  -8.83     &   16  &      29  \\
$\ge$~5001  &  &  5.7  &  -8.34     &    7   &     16  \\
     Whole      &&  3.4  &  -7.77      & 34   &    122   \\ 
\noalign{\smallskip}
\hline
\end{tabular}
\end{flushleft}
\end{table}

\section{Discussion} 

The objects presented in this paper have been divided into two operationally defined sub-samples: the ~$inner$~ sample and the ~$outer$~ sample. This distinction between the two sub-samples is of course arbitrary, and have no physical significance. However, there appears to be large differences between the average properties of the two samples, (Fig. 8 and 14). 

Very young "super star clusters", as the ones found in our inner sample have been found in a handful star forming dwarfs e.g. He2-10, NGC~1244, NGC~1569 and  NGC~1705 (for references see Sect. 1). 
However no intermediate age or old objects have been found in these galaxies, and sometimes there is  not enough information to constrain the ages e.g. when only one band-pass is observed as in the case of He2-10. In contrast, Thuan et al. (1996) found what appears to be old GCs in the blue compact dwarf Markarian~996.
Young to intermediate age globular cluster candidates have been found in large numbers in some merging galaxies e.g. 
NGC~3597, NGC~3921, NGC~4038/4039 and NGC~7252 (for references see Sect. 1). 
These are similar in luminosity and colour to the objects in our outer sample, (see Fig. 16 in Schweizer et al. 1996). For some galaxies, like NGC~7252, the luminosity function extends to much higher luminosities. This can be explained as a statistical effect since the number of objects is much greater and the host galaxy much more massive. Similar objects have also been found in kiloparsec scale circumnuclear rings in giant spiral galaxies (e.g. Barth et al. 1995), which may be related to bar instabilities and resonaces.

The inner sample is dominated by young (1 to 30 Myr) objects with masses ranging from $10^4$ to a few times $10^6$ ${\cal M_{\odot}}$ (for Salpeter and Miller-Scalo IMFs). Meurer et al. (1995) identified  9 of these in their study of this galaxy with the Faint Object Camera on board HST. 
These objects are young spatially extended star clusters with absolute magnitudes in V ranging from  $-15.5$ to $-9$. They are similar in properties to the super star clusters found in He2-10, NGC~1244, NGC~1705 and  NGC~1569, but the number of objects in ESO~338-IG04 is much greater, which could be a consequence of the higher starburst luminosity. The properties are fully consistent with young globular clusters. Among dwarfish galaxies, ESO~338-IG04 appears to be an unusual object in terms of the age span and the numbers of GCs.

\subsection{The luminosity function}

\paragraph{Outer sample}
In Fig.  8 we show the luminosity and  colour distributions for the outer sample. 
Since the clusters have a wide range of ages and there is strong luminosity evolution with age, the luminosity function (LF) will
change its shape as time goes by, even if no new clusters are added or no old ones disrupted.  
In Fig. 14 we show the luminosity function where all the absolute magnitudes have been transformed to an age of 9 Gyr. This was accomplished by using the model of photometric evolution and a Salpeter IMF (Sect. 5). This is more suitable for a direct comparison with galactic GCs. The result for the outer sample is a luminosity function not too different in shape from the Galactic one, but with the peak at higher luminosity, $M_v \sim -8.5$, which may partly be due to incompleteness. The Milky Way luminosity function peaks at $M_V=-7.2 \pm 0.2$ (Gnedin 1997). The number statistics are not good enough to distinguish between a power law and gaussian shape of the upper part of the luminosity function, and the apparent shape also depends on the chosen bin-size. A power law\footnote{The luminosity function expressed as a power law characterized by $\beta$, the power law index : $dN/dL \propto L^{-\beta}$} with power law index $\beta$~ in the range 1.5 to 2.0 is consistent with the data.  

\paragraph{Total sample}
The photometry of the total sample is of varying quality for sources with the same luminosity as has been discussed, and a few objects have been discarded due to their photometric uncertainties. Furthermore, due to severe crowding in the inner sample the completeness is bad at faint magnitudes and some objects may be unresolved multiple clusters. Therefore the luminosity function of the total sample is uncertain, especially at faint magnitudes. It is interesting to note that the peak in the $M_{v,9Gyr}$ distribution (Fig. 14)  now is quite close to the $M_V=-7.2 \pm 0.2$ peak in the luminosity function for Galactic GCs (Gnedin 1997).  

Van den Bergh (1995a) has argued  that the luminosity function of observed blue globular cluster candidates in the "Antennae" (Whitmore and Schweizer 1995) has the shape of a power-law, and is more similar to the galactic open cluster  than the galactic globular cluster luminosity function, which has a gaussian shape. He suggests that it is more likely that these are actually luminous open clusters. However, as was noted by Meurer (1995), the shape of the luminosity function may change as the objects fade and less massive clusters are preferably  disrupted. Furthermore as we have shown, since the clusters have different ages they will not fade equally much in a Hubble time. Many authors (e.g. Harris and Pudritz 1994, Elmegreen and Efremov 1997) have lately
questioned that the apparent gaussian luminosity function for galactic GCs reflects the initial distribution of proto-globular cluster masses. Transforming all objects to an equal age (9 Gyr) using the photometric evolutionary model, gives a luminosity distribution  similar to that  of Milky Way globular clusters (Fig. 14). However since the cut-off transformed luminosity depends on the present age of the object, the low luminosity end of the distribution is distorted. The number of objects is too small to discriminate between a power law and gaussian shape for the luminosity function. Both a gaussian and a power law with $\beta$ in the range  1.5 to 2.0 is consistent with the data.

\paragraph{Young objects ($\le 50$ ~Myr)}
Studying the luminosity distribution for young clusters(age $< 50$ Myr), the range covered in $M_v$ at 9 Gyr is ~$\sim -4 to -9.5$ with a peak at $M_v \sim -7$ and a median luminosity $<M_v>=-6.4$, for a Salpeter IMF, while for the Miller-Scalo IMF the peak luminosity is similar and  $<M_v>=-7.1$, cf. Table 4. Apparently the luminosity function for young objects is shifted to lower luminosity as compared to the older (age $> 50$ Myr) objects and the Milky Way GCs. However the range of luminosities is consistent with Galactic GCs. This may indicate that the young sample is contaminated with associations not deemed to become GCs. A more interesting interpretation is that all these are GCs in their initial configuration. Disruptive processes will affect, primarily, low mass systems, which would result in a peak in the luminosity function that is shifted upwards with age. Assuming the final luminosity function to be gaussian, with peak luminosity ($M_V=-7.2 \pm 0.2$) and dispersion ($\sigma_L=1.3$) similar to the situation in the Milky Way ($M_V=-7.2\pm 0.2, ~\sigma_L=1.4$, Gnedin 1997), Virgo dwarf ellipticals ($M_V=-7.0, ~\sigma_L=1.15$, Durrell et al. 1996) and nearby galaxies ($M_V=-7.1, ~\sigma_L=1.3$, Harris 1991), we expect only ~$33_{-10}^{+20} \%$~ of the young clusters to survive long enough to ever become old globular clusters, when compared to the $M_v$ distribution at 9 Gyr. The corresponding value for a Miller-Scalo IMF is ~$65_{-20}^{+35} \%$. These values should be viewed critically since they are based on low number statistics and lowering the Hubble constant will increase the expected number of surviving sources. The lower peak luminosity of the young objects may be coincidental and unrelated to destruction processes. We note that the metal rich subpopulation (which probably is younger than the metal poor subpopulation) of GCs in ellipticals have a
lower peak luminosity  than the metal poor subpopulation (Forbes et al. 1997).

In Table 4 we show the median mass and the median absolute magnitude transformed to 9 Gyr for the total sample as a function of age for the Salpeter and Miller-Scalo IMFs. In addition, for each interval of modelled ages we tabulate  $n$ the number of objects, and $n_{<-8.5}$ the number of objects with $M_{v,9Gyr} < -8.5$, i.e. at least one $\sigma$ brighter than the assumed peak. This last quantity was multiplied with 6.25 to estimate the number of surviving young clusters (assuming the final LF to be gaussian with a peak at ~$M_v =-7.2$, and $\sigma_L =1.3$). 

For each age interval, the luminosity distribution is consistent with a power law luminosity function with $\beta$ in the range 1.5 to 2.0. There is no clear relation between $\beta$ and the age of the objects. There is a weak dependence of $\beta$ on the adopted IMF: The Salpeter and Miller-Scalo IMFs give best fitting $\beta$ in the range 1.4 to 1.9, while the flat and steep IMFs give smaller and larger $\beta~$ values, respectively.

\subsection{Cluster sizes and kinematics}

In section 4.4 we showed that $\ge 95 \%$ of the objects are spatially resolved. The clusters in our sample have estimated sizes which are larger than those of Galactic GCs. However, since even HST under-samples a GC at these distances the estimated sizes should be regarded as upper limits (Schweizer et al. 1996). The uncertainties are further discussed in sect. 4.4. 
A strong support for the GC interpretation is that the estimated core and affective radii falls within the range observed for galactic GCs.

The typical core crossing time for a star within our GC candidates is 1 Myr, assuming a core radius of 3 pc and a velocity dispersion of 3 km/s.  If our objects were not gravitationally bound they would dissolve in a few crossing times. This means all objects older than 10 Myr, i.e. 80 percent of the whole sample, must be gravitationally bound. Thus it seems more likely that although some objects may consist of multiple associations along the line of sight, most inner objects must be newly formed globular cluster adepts. If they will eventually become old GCs, like the ones we see in the outer sample and the Milky Way,  depends on their dynamical interaction with the rest of the matter in the centre of the galaxy. If they have developed deep potential wells, they will have a high probability of surviving the starburst environment and make their way to become globular clusters as those found in the outer sample. The fact that we have found intermediate age GCs in the outer sample, supports that globular clusters       can form in large numbers at low redshift. 
Since ESO~338-IG04 is a galaxy with rather low mass it is probably also a more favourable place for globular clusters to survive disruption

\subsection{Cluster ages}

The results from the age modelling  are robust to changes in IMF and photometric errors as was shown in Sect 5. 
Since the galaxy and an old globular clusters population was formed $\sim$10 Gyr ago, this galaxy has experienced  additional major GC formation events. The last one is apparently still going on. Major formation events appear to have  occurred 0.1, 0.6, 2.5 and 5 Gyr ago. The strongest peak is found at ~$2.5 \pm 1$~Gyr with 20 detected GCs. It is possible that there was a more or less continuos GC production during the epoch 2.5 to 5 Gyrs ago rather than separate peaks, or that the two peaks rather reflects a metallicity bimodality among these objects. Either altermative is interesting, since a possible metallicity bimodality among objects of similar age, signifies different formation histories. In general, including also uncertainties in metallicity and internal extinction we expect the absolute ages to be accurate to within 50 \%. At present, we can only say that there are GC populations present with an ages ~$\le 30$~Myr, ~$100 \pm 40$~Myr, ~$600 \pm 300$~Myr,  ~$3.5 \pm 2 $~ Gyr and ~$11 \pm 4$~ Gyr (the quoted uncertainties correspond to the maximun expected deviation). 
 
Table 4 lists the number of objects in different age intervalls. However, as we have seen, the mass detection limit is age dependent. To estimate the total number of GCs in each age intervall (including objects below the detection limit) we use the number of objects with $M_{v,9Gyr} < -8.5$ (which should comprise 16 \% of the total number of GCs at each age on the simplified assumption of a universal and gaussian LF for GCs, cf Sect. 6.6). Then we see that slightly more than half the total number of GCs belong to the 2.5-5 Gyr population. About one fifth of the GCs are older than 5 Gyr, and one fourth younger than 1 Gyr. Thus the apparent situation in Fig. 14, that very young objects dominate the total number of GCs, is not real, but created by the age dependent mass detection limit.
However, over the last 1 Gyr, the total number of GCs formed is  comparable to the any of the 2.5 or 5 Gyr peaks. We do not know how extended in time these events were. Thus averaged over 1 Gyr, the 2.5 to 5 Gyr cluster formation episode(s) cannot be said with confidence to have been much  stronger than the present one.  

The peaks at 2.5 and 5 Gyr ago corresponds to redshifts of $z\approx 0.3$ and $z\approx 0.7$ respectively for the used value of $H_0 = 75 ~{\rm km~s^{-1} ~Mpc^{-1}}$, only weakly dependent on the value of $\Omega$.~
Some investigations  suggest that the "Faint blue galaxies" phenomenon, i.e. the excess in the deep  blue galaxy counts, have its origin in star bursting dwarfs at these redshifts (See e.g. Ellis 1997 for a review). We have found about five "real" peaks in the age distribution.  A GC formation event must be associated with a star burst, as is the case presently. Thus this galaxy should have experienced about five bursts of star formation in its lifetime. Thus age dating of GC formation events is an efficient way of studying the violent star formation history in this, and similar, galaxies.

\subsection{Cluster masses}

The estimated masses of our GC candidates are similar to those of galactic GCs. In Table 4 the median masses of the GC candidates are tabulated for different age intervals. The estimated mass is quite sensitive to the adopted IMF. At least for old clusters, the Miller-Scalo IMF should give realistic mass estimates since it reproduces the observed mass luminosity ratio of Galactic GCs (cf. Sect 5.4.1). Since there is no evidence that the IMF is different for  GCs in ESO~338-IG04 and the Milky Way, the Miller-Scalo IMF is the natural choice. 
The uncertainty in metallicity of the objects have no significant effect on the estimated masses, as compared to the other uncertainties.  

The photometric mass correlates with the fitted age in such a way that older clusters on average are more massive (Table 4, Fig. 15). This is a simple consequence of luminosity evolution  combined with a larger number of old clusters compared to young ones. Since a GC becomes fainter with time we will be able to detect clusters with lower mass for young ages than for old. Moreover, if the total number of old globular clusters is larger than the number of young ones, we expect the most massive old clusters to be heavier than the most massive young clusters.  

The highest mass is estimated for object no.  34 (Table 1) in the outer sample, which 
has a photometric mass of $ \ge 10^{7}  {\cal M_{\odot}}$ and an age of 2.5 Gyr for the both the Salpeter and Miller-Scalo IMF. The other
two IMFs give the same age and an even higher modelled mass. Thus, irrespective of the uncertainties
in the IMF, it appears to be a very massive globular cluster. Several almost as massive objects are found in both the inner and outer sample.

\subsection{The stellar IMF}

In Sect. 5 it was shown that there is strong support for the Miller-Scalo IMF, since it best reproduced the $M/L$ values of old galactic GCs. The flat and steep IMFs are ruled out by producing $M/L$ ratios an order of magnitude too large. 
On the other hand, the Miller-Scalo IMF have problems reproducing the $(v-i)$ colours of the bluest objects. 
Does this suggest that the IMF has changed or that we should take the Salpeter IMF which provides a fair compromise in these two respects?

There is no a priori reasons to assume that the IMF should change with time. There have been claims of IMF deficient in low mass stars in some starburst galaxies, but the results are inconclusive. Old GCs must be fossil starburst clusters, due to their small physical size. Recent HST results give support for a Miller-Scalo like mass function in the low mass region from direct star counts  in old Galactic GCs (e.g. de Marchi and Paresce, 1997). There seems to be no reason why the IMF should be different in the present starburst clusters than the previous ones. One alternative is however that the IMF is invariant but that the mass function evolves with time due to dynamical effects.

Once formed, a globular cluster continuously evolves. Stars, mainly of low mass, are  evaporated due to stellar collisions. In effect, the mass function is modified, not only by the death of massive stars, but also the low mass region is affected through dynamical processes. When supernovae explode, debris is ejected with high velocity and some material will leave the cluster. Also, the kinetic energy received by the remnant will give this a higher than average probability of escape.
Thus, we expect a globular cluster to continuously decrease in mass, and the stellar mass function may change in  shape.

This means that the IMF may be of Salpeter shape, while subsequent dynamical evolution may bring the low mass region of the mass function into close resemblance with the Miller-Scalo IMF. With such an  evolution of the stellar mass function in GCs, both the bluest colours and the $M/L$ ratios can be reproduced.

\subsection{Specific frequency of globular clusters}

The specific frequency of globular clusters $S_N$, is a measure of the relative richness of globular clusters of a galaxy. It is defined as $S_N = N_{clust}~ \times~10^{0.4(15+M_V)} $ ~(van den Bergh, 1985), where $M_V$ is the absolute magnitude of the host galaxy (for ESO~338-IG04 ~$M_v = -19.3$). In general, $S_N$ is a strong function of galaxy type with ellipticals having the highest specific frequencies, $S_N \approx 5$ while irregulars and late type spirals have $S_N \approx 1 $  (Harris, 1991; van den Bergh, 1995). This is a classical objection against the scenario that the majority of elliptical galaxies formed by merging late type galaxies. We have detected more than 120 globular cluster candidates, and even more are expected to hide beyond the detection limit. If we assume that all the 120 objects we are discussing in this paper are globular clusters we get $S_N \sim 2$. A lower conservative boundary on $S_N$ can be derived using only the objects in the inner and outer sample with age $\ge 50$Myr, which totals  65. If these were anything but GCs, they should have dissolved by now.  As was discussed is sect. 4.5 the expected number of interlopers, i.e. foreground stars and background galaxies, still contaminating our sample is expected to be six or less. Subtracting five from the number of old detected GCs we are left with 60.  This gives a lower limit of  $S_N \ge 1$. 

A more fair estimate of the richness of a globular cluster population is obtained by calculating the $T$-value (Zepf \& Ashman 1993) which relates the number of clusters to the mass of the host galaxy by also taking its $M/L_V$  into account. Beeing involved in a starburst, ESO~338-IG04 undoubtly has a low $M/L_V$ value, so by adopting $M/L_V = 2.0$ as suggested by Zepf \& Ashman (1993) for irregulars we will definetely not overestimate the $T$-value, but rather get a lower bound. For a total number of 120 GCs we then obtain $T \ge 13 $ which is very high even for giant elliptical standards. Counting only GCs older than 1 Gyr (which should have orbits favourable for non destruction) we still get 45 GCs and  $T \ge 5$, see Table 5. This is without any correction for objects below the detection limit. Thus, for beeing a late type galaxy, ESO~338-IG04 appears to be very rich in GCs.  

To estimate how many sources that may fall beyond the detection limit we assumed a gaussian luminosity function with a peak at $M_{V \rm peak}=-7.2 \pm 0.2$ and a one sigma width of the peak of $\sigma_L=1.3$ magnitudes, which is a compromise between values for the Milky Way (Gnedin 1997, Harris 1991) dwarfs (Durrell 1996) and nearby galaxies (Harris 1991). For a gaussian (normal) distribution only 16\% of the objects should have luminosities brighter than $M_{V {\rm peak}}-\sigma_L=-8.5$. Thus we can use $n_{< -8.5}$ to estimate the total number of GCs, including those behind the detection limit.  From Table 4 we see that in total 33 objects have  $M_{v,9Gyr} < -8.5$. Thus we expect the total number of globular clusters to be $\approx 205 \pm 30$, i.e. only $\sim 60\%$  of the GCs in  ESO~338-IG04 have been detected. Comparing the number of objects with $M_{v,9Gyr} < -8.5$ with the total number of objects in each age interval in Table 4, it is clear that all the undetected GCs are expected to be older than 50 Myr, and the majority (almost 90 \%) older than 1 Gyr. If some young objects will not make their way to become old GCs it has little effect on the total expected number of globular clusters, since almost all the objects brighter than  $M_v=-8.5$ are older than 50 Myr, and the majority (75 \%) older than 1 Gyr. Thus the total number of clusters is expected to be almost 200 anyway. 

The number of undetected clusters  depends on the assumed peak luminosity, the dispersion $\sigma_L$, and of course on errors in the adopted distance and the transformation of the luminosities to 9 Gyr. If, for example, $H_0$ would be 65 rather than the assumed value of 75 km~s$^{-1}$Mpc$^{-1}$, our globular clusters would become 0.3 magnitudes brighter and the total number of expected clusters would grow to more than 250. However, the host galaxy would become correspondingly brighter, almost cancelling the effect. With these uncertainties in mind we conclude that the total number of clusters is $ 200 \pm 50$, and  arrive at $S_N=4 \pm 1 $. This is a typical value for elliptical galaxies (Harris 1991). 
The corresponding $T$-value corrected for undetected GCs would however be very high ($ T \ge 20$). Even if counting only the expected number of old ($>1$~ Gyr) GCs we get  ~$ T \ge 17$.   

Since the host galaxy is involved in an active luminous starburst it is in a transient state where it is over-luminous as compared to galaxies with similar mass and size. In a Hubble time a starburst will fade 4-5 magnitudes (section 5), and about 3 magnitudes in 1 Gyr (Fig. 11). However, since there is an underlying  intermediate age component with lower luminosity and larger extent than the starburst, the fading of the galaxy as a whole would only be approximately 1.5 magnitudes in 1 Gyr as the starburst ceases. Thus, if the number of clusters is preserved the specific frequency would rise with a factor of $\sim$4 in one Gyr. During this period many young GCs maight be disrupted, but not enough to cancel the effect. Hence, in one Gyr, this galaxy will become very rich in GCs; even if only the detected old  ($> 1 $ Gyr) objects are counted $S_N \approx 3$. Including the expected number of old ($> 1 $ Gyr) objects behind the detection limit we expect 150 objects and $S_N\approx 11$. Taking also young ($< 1$~ Gyr) objects into account and assuming that they survive we expect 200 GCs which yields  $S_N\approx 15$. Thus irrespective of the uncertainties in the total number of clusters, when the present starburst terminates ESO~338-IG04 will become very rich in globular clusters within 1 Gyr. Values of $S_N \ge 10$  are only found in E and cD galaxies. It is interesting to note that cD galaxies also are believed to have a multiple merger history. We conclude that even conservative estimates on ~$S_N$~ and ~$T$~ yields very high values of the richness of the GC system in ESO~338-IG04.  
In Table 5 we present $S_N$ values for different cases described in this section.

\begin{table}
\caption[]{Specific frequency of globular clusters.  "$S_N$-now" is the present value of the specific frequency and "$S_N$-1Gyr" is the predicted specific frequency in 1 Gyr, taking the luminosity evolution of the galaxy into account and assuming that all GCs survive. To estimate the number of undetected sources, we assumed a gaussian luminosity function with a peak at $M_V=-7.2$, and a standard deviation of 1.3 magnitudes. The entry on the number of detected old GCs have been reduced with 5 to correct for possible interlopers. "T" is the specific frequency as defined by Zepf and Ashman (1993). For a discussion on the table entries, see text, Sect. 6.6.}
\begin{flushleft}
\begin{tabular}{lllll}
\noalign{\smallskip}
\hline\noalign{\smallskip}
~~~~~~ & $N_{clust}$ & $S_N$-now &  $S_N$-1Gyr & $T$\\
\noalign{\smallskip}\hline
\noalign{\smallskip}
GCs $> 50$ Myr, detected & $\ga$ 60      & $\approx$ 1 &  $\approx$  5 &$\ge 6$  \\
GCs $> 1$ Gyr, detected  & $\approx$ 45  & $\approx$ 1 &  $\approx$  3 &$\ge 5$  \\
GCs $< 50$ Myr, detected &           57                                          \\
GCs Total, detected      & $\approx$ 120 & $\approx$ 2 &  $\approx$  9 &$\ge 10$ \\
GCs $> 1$ Gyr, expected  & $\approx$ 150 & $\approx$ 3 &  $\approx$ 11 &$\ge 13$ \\
GCs Total, expected      & $\approx$ 200 & $\approx$ 4 &  $\approx$ 15 &$\ge 17$ \\
\noalign{\smallskip}
\hline
\end{tabular}
\end{flushleft}
\end{table}

\subsection{Formation of globular clusters}

The final question is: -- Why has this galaxy, at different epochs, been forming and probably still is forming, globular clusters in large numbers? The best guess at hand would be that this cluster formation is connected to an ongoing merger. This is supported by the velocity field (\"Ostlin et al. 1997a). An alternative explanation may be interaction with the 70 kpc distant confirmed companion galaxy. This galaxy is about 2 magnitudes fainter and also show signs of active star formation. It would be interesting to know if it possesses the same kind of clusters as ESO~338-IG04.
The formation of globular clusters requires a substantial mass of gas to be transformed into stars in a very limited region of space. The process of GC formation therefore requires both high gas densities and pressures, and high overall gas masses (Elmegreen and Efremov 1997). In addition, the timescale for bulding up the required gas masses and densities, must be shorter than the timescale for destruction of molecular clouds. These  conditions are fulfilled in mergers involving a late type galaxy. It is not certain at all, if a mere tidal interaction can trigger enough gas flows to create the necessary conditions. The velocity dispersion in the ionised gas in  ESO~338-IG04 is high,  ${\rm FWHM}\approx 100$ km/s (\"Ostlin et al. 1997a, Iye et al. 1987), indicating a high pressure and favourable conditions for GC formation.

It seems like objects similar to our young ones can form also in large (kiloparsec scale) cirumnuclear rings in giant spirals (Barth et al. 1995). The cluster formation in such environments is probably triggered by dynamical resonances enhancing the local gas density. Such mechanisms are lacking in dwarf galaxies, why mergers may be the only efficient way of GC formation. Once formed a GCs have better chances of surviving in a dwarf galaxy since the destruction mechanisms (tidal shocking, disk shocking, bulge shocking etc) are less efficient or absent in dwarfs.

The age distribution of the clusters indicates several different formation epochs have taken place, suggesting that the galaxy has been involved in a merger or strong interaction more than once since it and the original clusters were formed. The presently ongoing merger has distorted the velocity field enough that it is impossible to derive information on any previous merger. Earlier cluster formation epochs could possibly also be connected to a previous very close encounter between the galaxies that now are merging.
 
If we accept the merger scenario we then ask what kind of galaxies have been involved? A quite substantial population of old clusters suggest that one of the initial galaxies was an elliptical (dwarf) galaxy, while the existing large amounts of gas needs a late type galaxy for it's explanation. Of course it is also possible that the galaxy has merged with gas clouds rather than bona fide galaxies.  The specific frequency of GCs suggests that, if the present amount of gas can be consumed or expelled, this galaxy will probably evolve into an dwarf E/SO galaxy once the merger remnant relaxes. The event 2.5-5 Gyr ago must also have involved a gas rich dwarf in order to provide raw material for the large number of GCs formed then. An alternative explanation could be that the 2.5-5 Gyr population partly was "stolen" from another dwarf elliptical.

\section{Conclusions}

Multi-colour photometry with HST/WFPC2 of the metal poor blue compact galaxy ESO~338-IG04 (Tol~1924-416) has revealed a rich population of faint point-like sources in and surrounding the galaxy. Aperture photometry has been performed on these, and the photometric results have been transformed into ages and masses using a spectral evolutionary synthesis model. Special care was taken to assure that the sources are physically associated and not chance projected background or foreground sources. The results can be summarised as follows:

\begin{enumerate}

\item The objects discussed in this paper have absolute magnitudes, ~$M_v$, in the range -7 to -15 and $(v-i)$ colours ranging from less than -1 to almost 2.

\item The vast majority of the found objects are spatially resolved star clusters, physically associated with the galaxy. The outer objects follow the luminosity distribution of the galaxy closely (Fig. 9).

\item The number of interlopers, i.e. foreground stars, background galaxies and super giants in the target galaxy, in the presented sample is conservatively estimated to be less than ten objects. This leaves us with  more than 112 detected star clusters in  ESO~338-IG04.

\item Using the photometric evolution model we show that the objects, in general, are well fitted by  Salpeter and Miller-Scalo IMFs, and that the resulting age distribution is insensitive to the adopted IMF. The ages of the clusters range from a few Myr to more than 10 Gyr.

\item The objects have inferred masses  ranging from $10^4$ to more than $10^7 {\cal M_{\odot}}$. 

\item The above listed properties show that the objects are massive globular clusters of varying age.

\item There are several peaks in the globular cluster age distribution. This shows that there have been several globular cluster forming events in this galaxy. These appear to have occurred ~$\sim$ 11 Gyr ago, 2.5-5 Gyr ago, 600 Myr ago, 100 Myr ago and "now". The present event has lasted for a few times 10 Myrs. 

\item More than half of the total expected number of GCs have an age  in the range 2 to 5 Gyr. This peak consists of two subpopulations with either different age ($\Delta_{\rm AGE} \sim 2.5$~Gyr) or metallicity ($\Delta_{\rm [Fe/H]} \sim 1$~dex).

\item The specific frequency, $S_N$, is high in this galaxy. Taking into account that the fading starburst will decrease the luminosity of the galaxy by 1.5 magnitudes in 1 Gyr, this galaxy will be come very rich in globular clusters, even if the newly formed GCs don't survive. The predicted specific frequency of globular clusters is much larger than for late type galaxies and comparable to that for giant ellipticals.

\item We suggest that a merger is responsible for the current starburst and cluster formation. In view of the high specific frequency of old globular clusters suggests that an elliptical galaxy is a main ingredient in this system. To provide gas for star and cluster formation a gas rich dwarf must be the other main ingredient.  A major formation event 2.5-5 Gyr ago suggests that a merger involving one gas rich component also occurred at that time.

\item In view of the high specific frequency and the present merger, we suggest that this galaxy will evolve into a moderately luminous elliptical galaxy, unless disturbed by possible future interactions.

\item  This investigation for the first time gives strong support for a rich population both old and newly formed GCs in a metal poor blue compact galaxy. 

\item  Detection and dating of GCs in moderately distant BCGs is feasible using the HST and offers an efficient way of studying the violent star formation history, if also the effects of metallicity and extinction can be handled.

\end{enumerate}

\begin{acknowledgements}

Leif Festin is thanked for stimulating discussions on photometry and growth curves. Ib Koersner at the department for Radiation Sciences in Uppsala is thanked for his kind help with some early problems reading the tape. Kjell Olofsson is thanked for useful discussions and comments on the manuscript.We thank the referee Stephen Zepf for his comments that contributed to improve the quality of the paper. 
This work was  supported by the Swedish Natural Science Research Council and The Swedish National Space Board. This work was based on observations with the NASA/ESA {\it Hubble Space Telescope}, obtained at the Space Telescope Science Institute, which is operated by the association of Universities for Research in Astronomy, Inc., under NASA contract NAS5-26555.

\end{acknowledgements}

\end{document}